\documentstyle[prabib,aps,psfig,12pt]{revtex}

\newcommand{\al}{\alpha}
\newcommand{\be}{\beta}
\newcommand{\ga}{\gamma}
\newcommand{\la}{\lambda}
\newcommand{\de}{\delta}

\begin{document}

\title{Vacuum polarization of  massive scalar fields\\
in the spacetime of the electrically charged nonlinear 
black hole}

\author{Jerzy Matyjasek\thanks{Electronic Address:
matyjase@tytan.umcs.lublin.pl, jurek@iris.umcs.lublin.pl}}

\address{Institute of Physics, Maria Curie-Sk\l odowska University,\\
pl. Marii Curie - Sk\l odowskiej 1,\\
20-031 Lublin, Poland}

\maketitle

\begin{abstract}
\noindent
 The approximate renormalized stress-energy tensor of the quantized
 massive conformally coupled scalar field in the spacetime of
 electrically charged nonlinear black hole is constructed. It is
 achieved by functional differentiation of the lowest order of the
 DeWitt-Schwinger effective action involving coincidence limit of the
 Hadamard-Minakshisundaram-DeWitt-Seely coefficient $a_{3}.$  The
 result is compared with the analogous results derived for the
 Reissner-Nordstr\"om  black hole. It it shown that the most important
 differences occur in the vicinity of the event horizon of the black
 hole near the extremality limit.	 The structure  of the nonlinear
 black hole is briefly studied  by means of the Lambert functions.

\end{abstract}

\vskip 0.8cm \noindent {PACS numbers: 04.70.Dy, 04.62+v \\UMCS-TG-00-18}

\preprint{}
% \clearpage
 
 \section{introduction}
For the quantized massive fields in the large mass limit, i. e., when
the Compton length is much smaller than the characteristic radius of a
curvature,  the nonlocal contribution to the effective action can be
neglected, and the series expansion in $m^{-2}$ of the renormalized
effective action, $W_{R},$  may be easily constructed with the aid of
the DeWitt-Schwinger method~[1,2]. As the  renormalization prescription
requires absorbtion of the first three terms of the series  into the
quadratic classical gravitational action, the $n-{\rm th}$ term of
$W_{R}$ is proportional to the integrated coincidence limit of the
Hadamard-Minakshisundaram-DeWitt-Seely coefficient (HMDS) $a_{n+3}.$
Unfortunately,  the complexity of the coefficients $[a_{n}]$ rapidly
increases with $n,$ and consequently one expects that the
applicability of the series expansion  is confined to a first, perhaps
a second nonvanishing term.

Having constructed a first order $W_{R},$ the renormalized
stress-energy tensor (which is the most important characteristics of
the quantized field	in the curved spacetime) may be obtained  in a
standard way, i. e., by functional differentiating the constructed
effective action with respect to the metric tensor. This method has
been successfully applied in calculations of the approximate renormalized
stress-energy tensor of the quantized massive scalar, spinor, and
vector fields in the vacuum type-D geometries by Frolov and Zel'nikov~[3-7].

A different method, based on the WKB approximation of the solutions of
the massive scalar field equation in a general spherically-symmetric
spacetime, and summation thus obtained modes	by means of the Abel-Plana
formula, has been invented by Anderson, Hiscock, and Samuel and applied
in the context of the RN spacetime~[8]. Their method is
equivalent to the Schwinger - DeWitt expansion: to obtain the lowest
(i. e. $m^{-2}$) terms, one has to use sixth-order WKB approximation.
Moreover, numerical calculations reported in Ref.[8] confirmed that the
DeWitt-Schwinger method provide a good approximation of the
renormalized stress-energy tensor of the massive scalar field with
arbitrary curvature coupling as long as the mass of the field remains
sufficiently large. This approach and its modifications has been
employed in various contexts in Refs.[9-15].

Recently, extending the 	results of Frolov-Zel'nikov,  
we have constructed a general formula describing the
approximate renormalized $\langle T^{b}_{a}\rangle_{ren}$ of the
quantized massive scalar, spinor, and vector fields in arbitrary
spacetime. The results have been presented  for the 
class of geometries with vanishing curvature scalar, and subsequently
applied in the spacetime of the RN black hole
and in the spacetime that could be obtained	by expanding its near horizon
geometry into a whole manifold~[16].
Our formulas allow, in principle, to
determine the renormalized stress-energy tensor of the massive field
once the line element has been chosen, although the specific
calculations may be very tedious.
For the quantized massive
scalar field with arbitrary curvature coupling in the
RN spacetime we have reproduced the results of
Anderson, Hiscock, and Samuel; neutral  spinor and vector fields have
not been discussed earlier.

In this paper we shall extend the analyses of Ref.[16] to the general
geometry and construct the renormalized stress-energy tensor of the
massive quantized scalar field obeying the equation
\begin{equation}
\left( - \Box \,+\,\xi R\,+\,m^{2}\right)\phi\,=\,0,
\end{equation}
where $\xi $ is the coupling constant and $m$ is the mass of the
field. Since the background geometry is general, the most direct
approach is to use the first nonvanishing term of the renormalized
effective action. The advantage of this approach lies in the purely
geometric nature of the approximation that reflects its local nature.
Although the constructed result is rather complex, we shall present it
in its full length, because it provides the generic formula from which
the renormalized stress-energy tensor in some physically interesting
cases may be easily obtained. As the effective action of the quantized
massive scalar field differs form the analogous actions constructed
for fields of higher spins only by numerical coefficients, one can
generalize presented results to fields of other spins. It should be
emphasized however, that the method has obvious limitations, and, when
applied to  rapidly varying or strong gravitational fields it breaks
down. Moreover, its massless limit is contaminated by nonphysical divergences.

Our general formulas will be employed in the calculations of
$\langle T_{a}^{b}\rangle_{ren}$ in the geometry  of the electrically
charged black hole, being an exact solution of the coupled system of
the Einstein equations and the equations of the nonlinear
electrodynamics recently proposed by Ay\'on-Beato and Garc\'{\i}a (ABG) in
Ref.~\cite{Beato}, to which the reader is referred for physical
motivations and technical details. Their exact solution is characterized
by the electric charge, $e$ and the mass $M,$ and may be elegantly
expressed in terms of the	hyperbolic functions. 
An important and interesting feature of this solution is its regularity
as radial coordinate tends to zero.
We shall show that the structure of
horizons of the ABG solution may by studied by means of the Lambert
function~\cite{Lambert}, allowing analytical treatment of the vacuum
polarization effects on the event horizon. At large distances their
solution behaves as the RN solution. For small and
intermediate values of the ratio $|e|/M,$ the location of the event
horizon, $r_{+},$ is  close to the location of the event horizon
of the RN black hole; significant differences occur near the
extremality limit. It would be, therefore,  interesting to analyze how
the similarities of the line elements are reflected in the behavior of
the renormalized stress-energy tensors. 

The renormalized effective action of the massive scalar field involves
the terms that are proportional to the first and third power of $\xi
-1/6. $ As the curvature scalar of the RN spacetime vanishes, $\langle
T^{b}_{a}\rangle_{ren}$ of the massive scalar field naturally divides
into the part that describes pure conformal coupling and  an
additional local part that is multiplied by a factor $\xi -1/6.$ On
the other hand however, the curvature scalar of the ABG geomerty does not
vanish, and  the structure of the effective action indicates that the
renormalized stress-energy tensor of the massive scalar depends on the
constant $\xi $	in a more complicated way. Since the conformal
coupling leads to massive simplifications, one expects that the
similarities  in the renormalized $\langle
T^{b}_{a}\rangle_{ren}$ (if any) would appear  mainly in this case.

 %%%%%%%%%%%%%%%%%%%%%%
 \section{The renormalized stress-energy tensor of the quantized massive scalar
 field}
  
  The renormalized effective action constructed for the quantized scalar field
satisfying equation (1)	 is given by
\begin{equation}
W_{ren}\,=\,{1\over 32\pi^{2}\,m^{2}} 
\int  d^{4}x g^{1/2} \sum_{n=3}^{\infty}{(n-3)!\over (m^{2})^{n-2}}[a_{n}],
\end{equation}
where $[a_{n}]$ is the coincidence limit of the $n-$th HDSM
coefficient. The first three coefficients of the DeWitt-Schwinger
expansion, $a_{0},\,a_{1},\,{\rm and}\,a_{2}, $ which contribute to the
divergent part of the  action have to be absorbed in the
classical gravitational action by renormalization of the bare coupling
constant.

As the rigorous asymptotic analysis of  the fundamental solution is
restricted, in general, to the heat operator of a parabolic type, we
tacitly assume that all steps that are necessary in construction of
the first-order renormalized stress-energy tensor have been carried
out in an analytically continued Euclidean spacetime. The analytic
continuation to the physical space is performed at the last stage of
the calculations.

Calculation of the HDSM coefficients is an extremely
laborious task, and their exact form for $n \geq 5$ is unknown. 
The coefficient $[a_{2}],$ which is proportional to the trace anomaly
of the renormalized stress-energy tensor 
of the quantized, massless, and conformally invariant fields,
has been calculated by DeWitt~\cite{DeWitt2}.
The coincidence limit of the coefficient $a_{3}$ has been obtained by
Gilkey [19,20] whereas the coefficient $[a_{4}]$ has been  calculated by
Avramidi [21-24].

Restricting ourselves to the terms  proportional to $m^{-2},$
one has
 \begin{equation}
W_{ren}  \,=\,{1\over 32\pi^{2}\,m^{2}} 
\int  d^{4}x g^{1/2} [a_{3}],
\end{equation}
 with 
\begin{equation}
[a_{3}]\,=\,{b_{3}\over 7!} \,+\,{c_{3}\over 360}, 
\end{equation}
where
\begin{eqnarray}
\nonumber
b_{3}\,&=&\,{35\over 9} R^{3}\,+\,17  R_{;p}  R^{;p}\,-\, R_{q a;p}
R^{q a;q}\,-\,4 	 R_{q a ; p} R^{p a ; q} \nonumber \\
&+& 9  R_{q a b c;p} R^{q a b c;p}
\,+\,2 R \Box R \,+\,
18 \Box^{2} R\,-\,8 R_{p q} \Box R^{p q}	 
- {14\over 3} R R_{p q} R^{p q}
\nonumber \\
&+&24 R_{p q ; a}^{\ \ \ q} R^{p a} 
\,-\,
{208\over 9} R_{p q} R^{q a} R_{a}^{\ p} \,+\, 12 \Box R_{p q a b}
R^{p q a b}
+ {64 \over 3} R_{p q} R_{a b} R^{p a q b}
 \nonumber \\
&-&{16\over 3} R_{p q}
R^{p}_{~ a b c} R^{q a b c}	
\,+\,{80\over 9} R_{p q a b}
R_{c ~ d}^{~ p ~ a} R^{q c b d}	\,+\,{44 \over 9}
R_{p q a b} R_{c d}^{~ ~ p q} R^{a b c d}
\end{eqnarray}
and
\begin{eqnarray}
\nonumber
c_{3}\,&=&\,-(5 \xi - 30 \xi^{2} + 60 \xi^{3}) R^{3}\,-\,( 12 \xi - 30 \xi^{2}) R_{;p}
R^{;p}\,-\,(22 \xi - 60 \xi^{2}) R \Box R\\	 
&-&\,6 \xi \Box^{2} R\,-\,4 \xi  R_{p q} R^{p q}\,+\,2 \xi R R_{p q}
R^{p q}\,-\,2 \xi R R_{p q a b} R^{p q a b}.
\end{eqnarray}
Since the  coincidence limit of the coefficient $a_{4}$ is much more
complex one expects that using it in the calculations of the approximate
renormalized stress-energy would be a real challenge.  However, it still 
could by of use in the  simpler analyses of the field fluctuation, 
$\langle \phi^{2}\rangle_{ren}$.
Substituting (4-6) into (3), integrating by parts and 
making use of elementary  properties of the Riemann tensor,
one can reduce 	the number of terms in the renormalized 
effective action to ten~\cite{avra1}:

 \begin{eqnarray}
 \nonumber
 W^{(1)}_{ren}\,&=&\,{1\over 192 \pi^{2} m^{2}} \int d^{4}x g^{1/2}
 \left[ \left({1\over 2} \xi^{2} - {1\over 5} \xi + {1\over 56}\right) R 
 \Box R\,+\,{1\over 140} R_{p q} \Box R^{p q}\right. \nonumber \\
 &+&
 \left({1\over 6} - \xi\right)^{3} R^{3} 
 \,-\,{1\over 30}\left({1\over 6} - \xi\right) R R_{p q } R^{p q}
 \,+\,{1\over 30}\left({1\over 6} - \xi\right) R R_{p q a b} R^{p q a b}
\nonumber \\ 
 &-& {8\over 945} R^{p}_{q} R^{q}_{a} R^{a}_{p}
 \,+\,{2\over 315} R^{p q}
 R_{a b} R^{a ~ b}_{~ p ~ q} 
\,+\,
\,{1\over 1260} R_{p q} R^{p}_{~ c a b} R^{q c a b}
\nonumber \\
&+&
\left.{17\over 7560} {R_{a b}}^{p q} {R_{p q}}^{c d}
{R_{ c d}}^{a b} \,
 \,-\,{1\over 270}
R^{a ~ b}_{~ p ~ q}
 R^{p ~ q}_{~ c ~ d} R^{c ~ d}_{~ a ~ b}\right] \nonumber \\
 &=&{1\over 192 \pi^{2} m^{2}}\sum_{i = 1}^{10} \al_{i} W_{i},
\end{eqnarray}
where $\al_{i}$ are numerical coefficients that stand in front of the geometrical 
terms.

 The renormalized stress-energy tensor is given by the standard relation
\begin{equation}
{2\over g^{1/2}}{\delta\over \delta g_{a b}} W^{(1)}_{ren}\,=\, 
\langle T^{a b}\rangle_{ren}.
\end{equation}
Functionally differentiating the renormalized effective action with
respect to the metric tensor, performing simplifications and  necessary 
symmetrizations, after rather long calculations, one has
%%%%%%%%%%%%%%%%%%%%%%%%%%%%%%%%
%%%%%%%%% W_{1} %%%%%%%%%%%%%%%%
\begin{eqnarray}
{1\over g^{1/2}}{\delta\over \delta g_{m n}}W_{1}\,&=&\,
R^{; m} R^{; n}\,+\,(\Box R)^{; m n}\,+\,(\Box R)^{; n m}\nonumber \\
&-&
{1\over 2} R_{; p}R^{; p} g^{m n}
\,-\,2 \Box^{2} R g^{m n}\,-\,2 \Box R R^{m n} ,
\end{eqnarray}
%%%%%%%%% W_{2} %%%%%%%%%%%%%%%% 
\begin{eqnarray}
{1\over g^{1/2}}{\delta\over \delta g_{m n}}W_{2}\,&=&\,
R_{p q }^{\ \ ; m} R^{p q ; n}\,-\, R_{p q  }^{\ \ ; n} R^{p m ; q}\,-\,
R_{p q  }^{\ \ ; m} R^{p n ; q}	 \nonumber \\
&+&
R_{p q }^{\ \ ; p} R^{q m ; n} 
\,+\,R_{p q  }^{\ \ ; p} R^{q n ; m}\,+\, (\Box R_{p}^{\ m})^{; n p}
\nonumber \\
&+&
(\Box R_{p}^{\ n})^{; m p}\,-\,\Box^{2} R^{m n}
\,-\,
{1\over 2} R_{p q ; r} R^{p q ; r}g^{m n}
\nonumber \\
&-&
(\Box R_{p q})^{; q p} g^{m n}\,+\,R_{p \ \ \ q}^{\ m ; n} R^{p q}
\,+\,
 R_{p \ \ \ q}^{\ n ; m} R^{p q} \nonumber \\
 & -&
 R_{p q}^{\ \ ; n q}R^{p m}\,-\,
 \Box R_{p}^{\ n} R^{p m}
 \,-\, \Box R_{p}^{\ m} R^{p n}\nonumber \\
 &-&
 R_{p q}^{\ \ ; m q} R^{p n} ,
 \end{eqnarray}
%%%%%%%%%%% W_{3} %%%%%%%%%%%%%%
%%%%%%%%%%%%%%%%%%%%%%%%%%%%%%%%%
 \begin{eqnarray}
 {1\over g^{1/2}}{\delta\over \delta g_{m n}}W_{3}\,&=&\,
 6 R^{; m} R^{; n} \,+\, 6 R R^{; m n}\,+\, {1\over 2} R^{3} g^{m n}
 \nonumber \\
 &-&
 6 R_{; p} R^{; p} g^{m n}
 \,-\, 6 R \Box R g^{m n}\,-\, 3 R^{2} R^{m n},
 \end{eqnarray}
%%%%%%%%%%%%%%%%%%%%%%%%%%%%%%%%
%%%%%%%%%%% W_{4} %%%%%%%%%%%%%%
 \begin{eqnarray}
{1\over g^{1/2}}{\delta\over \delta g_{m n}}W_{4}\,&=&\,
R^{; n} R_{p}^{m ; p}\,+\,R^{; m} R_{p}^{n ; p}	\,+\,2 R_{p q}^{\ \ ; n} R^{p q ; m}
\nonumber \\
&+& 
R_{; p} R^{p m ; n} 
\,+\,
R_{; p} R^{p n ; m} \,-\,
2 R_{; p} R^{m n ; p}
\nonumber \\
&+&
R R_{p}^{\ m ; n p}\,+\,R R_{p}^{\ n ; m p} 
\,-\,
R \Box R^{m n}
\nonumber \\
&-&
2 R_{; p} R_{q}^{\ p ; q} g^{m n}\,-\,
2 R_{p q ; r} R^{p q ; r} g^{m n}\,-\,R R_{p q}^{\ \ ; q p} g^{m n}	\nonumber \\
&+&
R_{p q}^{\ \ ; m n} R^{p q}\,+\,R_{p q}^{\ \ ; n m} R^{p q}	\,+\,
R_{; p q} R^{p q} g^{m n}
\nonumber \\
&-& 2 \Box R_{p q} R^{p q} g^{m n}
\,+\,
{1\over 2} R R_{p q} R^{p q} g^{m n}\,+\,R_{; p}^{\ \ n} R^{p m}
\nonumber \\
&-&
2 R R_{p}^{\ n} R^{p m}	\,+\,R_{; p}^{\ \ m} R^{p n} 
\,-\,
\Box R R^{m n}
\nonumber \\
&-&
R_{p q} R^{p q} R^{m n} ,
\end{eqnarray}
%%%%%%%%%%%%%%%%%%%%%%%%%%%%%%%%
%%%%%%%%%%% W_{5} %%%%%%%%%%%%%%
\begin{eqnarray}
{1\over g^{1/2}}{\delta\over \delta g_{m n}}W_{5}\,&=&\,
-4 R_{; p} R_{q}^{\ m p n ; q}\,-\,4 R_{; p} R_{q}^{\ n p m ; q}\,+\,
2 R_{p q r s}^{\ \ \ \ ; n} R^{p  q r s ; m}
\nonumber \\
&-& 
2 R R_{p \ q}^{\ m \ n ; q p }
\,-\, 2 R R_{p \ q}^{\ n \ m ; q p } \,-\,2 R_{p q r s ; t} R^{p q r s ; t} g^{m n}
\nonumber \\
&+& 
R_{p q r s}^{\ \ \ \ ; m n } R^{ p q r s}\,+\, 
R_{p q r s}^{\ \ \ \ ; n  m } R^{ p q r s}
\,-\,
2 \Box R_{p q r s} R^{p q r s}g^{m n}
\nonumber \\
&+&
{1\over 2} R R_{p q r s} R^{p q r s} g^{m n}\,-\,
R^{m n} R_{p q r s} R^{p q r s}\,-\,  
2 R R_{p q r}^{\ \ \ n} R^{p q r m} \nonumber \\
&-&
2 R_{; p q} R^{p m q n}\,-\,2 R_{; p q}R^{p n q m} ,
\end{eqnarray}
%%%%%%%%%%%%%%%%%%%%%%%%%%%%%%%%
%%%%%%%%%%% W_{6} %%%%%%%%%%%%%%
 \begin{eqnarray}
{1\over g^{1/2}}{\delta\over \delta g_{m n}}W_{6}\,&=&\,
{3\over 2} R_{p q}^{\ \ ; n}  R^{p m ; q}\,+\, {3\over 2} R_{p q}^{\ \ ; m}	R^{p n ; q}
\,-\, 3 R_{p \ ; q}^{\ m} R^{p n ; q}
\nonumber \\
&+&
{3 \over 2} R_{p q}^{\ \ ; p} R^{q m ; n}
\,+\, {3\over 2} R_{p q}^{\ \ ; p} R^{q n ; m}\,-\,
{3\over 2} R_{p q}^{\ \ ; p} R_{r}^{ \ q ; r} g^{m n}
\nonumber \\
&-&
{3\over 2} R_{p q ;r} R^{ p r ; q} g^{m n}\,+\,{3\over 2}R_{p \ \ \ q}^{\ m ; n} R^{p q}
\,+\,
{3\over 2} R_{p \ \ \ q}^{\ n ; m} R^{p q}
\nonumber \\
&-&
{3\over 2} R_{p q ; r}^{\ \ \ \ q} R^{p r} g^{m n}\,+\,
{3\over 2} R_{p q}^{\ \ ; n q} R^{p m}
\,-\,
{3\over 2} \Box R_{p}^{\ n} R^{p m}	\nonumber \\
&+&
{3\over 2} R_{p q}^{\ \ ; m q} R^{p n}\,-\, {3\over 2} \Box R_{p}^{\ m} R^{p n}
\,-\,
{3\over 2} R_{p q \ \ r}^{\ \ ; p} R^{q r} g^{m n}
\nonumber \\
&+& R_{p q} R_{r}^{\ p} R^{q r}
\,-\, 3 R_{p q} R^{p m} R^{q n} ,
\end{eqnarray}
%%%%%%%%%%%%%%%%%%%%%%%%%%%%%%%%
%%%%%%%%%%%%% W_{7} %%%%%%%%%%%%
\begin{eqnarray}
{1\over g^{1/2}}{\delta\over \delta g_{m n}}W_{7}\,&=&\,
R_{p}^{\ m ; p} R_{q}^{\ n ; q}\,+\,R_{p \ ; q}^{\ n} R^{q m ; p}\,-\,
2 R_{p q}^{\ \ ; p} R^{m n ; q}
\nonumber \\
&-&
R_{p q}^{\ \ ; n} R_{r}^{\ p q m ; r}
\,-\,
R_{p q}^{\ \ ; m} R_{r}^{\ p q n ; r}\,+\,R_{p q ; r} R^{p r q m ; n}
\nonumber \\
&+&
R_{p q ; r} R^{p r q n ; m}\,-\,2 R_{p q ; r}R^{p m q n ; r}
\,+\,
2 R_{p q ; r} R_{s}^{\ p q r ; s} g^{m n}
 \nonumber \\
&-&
R^{m n}_{\ \ ; p q} R^{p q}\,+\,
R_{p \ q r}^{\ m \ \ ; n r} R^{p q}\,-\,\Box R_{p \ q}^{\ m \ n} R^{p q}
\nonumber \\
&+&
R_{p \ q r} ^{\ n \ \ ; m r} R^{p q}\,-\,R_{p q r s}^{\ \ \ \ ; s q} R^{p r} g^{m n}\,+\,
{1\over 2} R_{p \ ; q}^{\ n \ \ p} R^{q m}
\nonumber \\
&+&
{1\over 2} R_{p \ \ \ q}^{\ n ; p }R^{q m}
\,+\,
{1\over 2} R_{p \ ; q}^{\ m \ \ q} R^{q n}\,+\,
{1\over 2} R_{p \ \ \ q}^{\ m ; p} R^{q n}	
\nonumber \\
&-& 
R_{p q}^{\ \ ; q p} R^{m n}
\,+\,
{1\over 2} R_{p q} R_{r s} R^{p r q s}
\,+\,
R_{p q \ \ r}^{\ \ ; n}R^{p r q m}
\nonumber \\
&-&
{3\over 2} R_{p q} R_{r}^{\ n} R^{p r q m}\,+\, R_{p q \ \ r}^{\ \ ; m}	R^{p r q n}\,-\,
{3\over 2} R_{p q} R_{r}^{\ m} R^{p r q n}
\nonumber \\
&-& 
R_{p q ; r s} R^{p s q r} g^{m n}\,-\,
\Box R_{p q} R^{p m q n} ,
\end{eqnarray}
%%%%%%%%%%%%%%%%%%%%%%%%%%%%%%%%
%%%%%%%%%%%%% W_{8} %%%%%%%%%%%%
\begin{eqnarray}
{1\over g^{1/2}}{\delta\over \delta g_{m n}}W_{8}\,&=&\,
-2 R_{p \ ; q}^{ \ m} R_{r}^{\ p q n ; r} \,+\,2 R_{p \ ; q}^{\ m} R_{r}^{\ n p q ; r}
\,-\, 2 R_{p q}^{\ \ ; p} R_{r}^{\ n q m ; r}
\nonumber \\
&+&
R_{p q r s}^{\ \ \ \ ; n} R^{p q r m ; s}
\,-\,
R_{p q r \ ; s}^{\ \ \ m}	R^{p q r n ; s}\,-\,2 R_{p q ;r} R^{p m r n ; q}
\nonumber \\
&-&
R_{p q r s}^{\ \ \ \ ; p} R^{q m r s ; n}\,-\,
{1\over 2} R_{p q r s}^{\ \ \ \ ;p} R_{t}^{\ q r s ; t} g^{m n}
\,-\,
{1\over 2}R_{p q r s ; t} R^{p q r t ; s} g^{m n}
\nonumber \\
&-&
2 R_{p \ q \ \ \ r}^{\ n \ m ; p} R^{q r} \,-\,
2 R_{p q r}^{\ \ \ n ; r p} R^{q m}\,+\,2 R_{p \ ; q r}^{\ m} R^{p r q n}
\nonumber \\
&+&
R_{p \ q r \ \ s}^{\ m \ \ ; n} R^{p s q r}\,+\, R_{p}^{\ m} R_{q r s}^{\ \ \ n} R^{p s q r}
\,-\,{1\over 2} R_{p q r s ; t}^{\ \ \ \ \ \ q} R^{p t r s} g^{m n}
\nonumber \\
&-&
{1\over 2} \Box R_{p \ q r }^{\ n} R^{p m q r}
\,+\, 
R_{p q r s}^{\ \ \ \ ; n q} R^{p m r s}\,-\,{1\over 2} \Box R_{p \ q r}^{\ m} R^{p n q r}
\nonumber \\
&+& 
2 R_{p q} R_{r \ s}^{\ p \ m } R^{q r s n}\,+\,
{1\over 2} R_{p q r s \ \ t}^{\ \ \ \ ; p} R^{q t r s} g^{m n}
\,-\,
{1\over 2} R_{p q} R_{r s t}^{\ \ \ p} R^{q t r s} g^{m n}
 \nonumber \\
&-&	2
R_{p q ; r}^{\ \ \ \ p}
R^{q m r n}	,
\end{eqnarray}
%%%%%%%%%%%%%%%%%%%%%%%%%%%%%%%%%
%%%%%%%%%%%% W_{9} %%%%%%%%%%%%%%
\begin{eqnarray}
{1\over g^{1/2}}{\delta\over \delta g_{m n}}W_{9}\,&=&\,
- 6 R_{p q r}^{\ \ \ n ; r} R_{s}^{\ m p q ; s}\,-\,
6 R_{p q r \ ; s}^{\ \ \ n} R^{p q s m ; r}\,-\,
3 R_{p q r}^{\ \ \ m} R_{s t}^{\ \ r n} R^{p q s t}
\nonumber \\
&-&
3 R_{p q r \ ; s}^{\ \ \ n \ \ r}R^{p q s m}
\,-\,
3 R_{p q r \ ; s}^{\ \ \ m \ \ r} R^{p q s n}\,-\,
3 R_{p \ q r \ \ s}^{\ n \ \ ; p} R^{qr s m}
\nonumber \\
&-&
3 R_{p \ q r \ \ s}^{\ m \ \ ; p}R^{qr s n}\,+\,
{1\over 2} R_{p q r s} R_{t u}^{\ \ p q} R^{r s t u} ,
\end{eqnarray}
and
%%%%%%%%%%%%%%%%%%%%%%%%%%%%%%%%%
%%%%%%%%%%%%% W_{10} %%%%%%%%%%%%
\begin{eqnarray}
{1\over g^{1/2}}{\delta\over \delta g_{m n}}W_{10}\,&=&\,
3 R_{p q r}^{\ \ \ m ; p} R_{s}^{\ r q n ; s}\,+\,
3 R_{p q r \ ; s}^{\ \ \ n} R^{p m r s ; q}\,+\,
3 R_{p q r s}^{\ \ \ \ ; p} R^{q m r n; s}
\nonumber \\
&+&
3 R_{p q r s}^{\ \ \ \ ; p} R^{q n r m ; s}
\,-\,{3\over 2} R_{p \ q \ ; r s}^{\ m \ n} R^{p s q r}\,-\,
{3\over 2} R_{p \ q \ ; r s}^{\ n \ m} R^{p s q r}
\nonumber \\
&+&
{3\over 2} R_{p \ q r \ \ s}^{\ n \ \ ; r} R^{p s q m}\,+\,
{3\over 2} R_{p \ q r \ \ s}^{\ m \ \ ; r} R^{p s q n}
\,-\,
3 R_{p q r}^{\ \ \ m} R_{s \ t}^{\ q \ n} R^{p s r  t}
\nonumber \\
&+&
{3\over 2} R_{p q r \ \ ; s}^{\ \ \ n \ \ q}R^{p m r s}\,-\,
{3\over 2} R_{p q r s}^{\ \ \ \ ; s q} R^{p m r n}\,+\,
{3\over 2} R_{p q r \ ; s}^{\ \ \ m \ \ q} R^{p n r s}
\nonumber \\
&-&
{3 \over 2} R_{p q r s}^{\ \ \ \; s q} R^{p n r m}\,+\,
{1\over 2} R_{p q r s} R_{t \ u}^{\ p \ r} R^{q t s u} .
\end{eqnarray}
%%%%%%%%%%%%%%%%%%%%% Koniec wyrazen geometrycznych.%%%%%%%%%%%%%%%%
%%%%%%%%%%%%%%%%%%%%%%%%%%%%%%%%%%%%%%%%%%%%%%%%%%%%%%%%%%%%%%%%%%%%
As there are numerous identities involving the Riemann tensor, its
covariant derivatives and contractions, the form of $\langle
T_{b}^{a}\rangle_{ren}$ is, of course, not unique and depends on
adopted simplification strategies. 
Here we presented our results in the form that we
have found useful in the further calculations.
It should be noted that the resulting
renormalized stress-tensor of the massive scalar field depends on the
coupling constant in a complicated way, and in a general spacetime
it divides naturally into four terms
\begin{equation}
 \langle T_{a}^{b}\rangle_{ren}\,=\,{1\over 30}\left(\xi - {1\over 6}\right)
{T^{(1)}}_{a}^{b}\,+\,
{1\over 2}\left(\xi - {1\over 6}\right)^{2} {T^{(2)}}_{a}^{b}\,+\,
\left(\xi - {1\over 6}\right)^{3} {T^{(3)}}_{a}^{b}
\,+\, {T^{(4)}}_{a}^{b},
\end{equation}
where

\begin{equation}
 T^{(1)a b}\,=\,{1\over 96 \pi^{2}m^{2}} 
 g^{-1/2}{\delta\over \delta g_{a b}}(W_{5} - W_{4}),
\end{equation}

\begin{equation}
T^{(2)a b}\,=\,{1\over 96 \pi^{2} m^{2}}g^{-1/2}{\delta\over \delta g_{a b}}W_{1},
\end{equation}

\begin{equation}
T^{(3)a b}\,=\,{1\over 96\pi^{2} m^{2}}g^{-1/2}{\delta\over \delta g_{a b}}W_{3},
\end{equation}
\begin{equation}
T^{(4) a b}\,=\,{1\over 96 \pi^{2} m^{2}}\,g^{-1/2}\left(
\beta {\de \over \de g_{a b}} W_{1}\,+\,
\al_{2}{\delta\over \delta g_{a b}}W_{2}\,+\,  
\sum_{i=6}^{10} \al_{i}{\delta\over \delta g_{a b}}W_{i}\right),
\end{equation}
and
\begin{equation}
\be\,=\,{1\over 252}\,-\,{\xi\over 30}.
\end{equation}
Inspection of eqs. (9-18) shows that variational derivatives of
$W_{1}$ and $W_{3},$ with respect to the metric tensor vanish in $R =0
$ geometries, and, additionally,  that of $W_{2},$  $W_{4},$ $W_{6},$
and $W_{7}$ vanish for the Ricci-flat geometries. Moreover, one has
important simplifications of the general stress-energy tensor for the
conformally coupled massive fields as there is no need to compute
$T^{(1)ab},$ $T^{(2)ab},$  and $T^{(3)ab}.$ Finally we observe  that
the analogous expression of the stres-energy tensor of the quantized
massive spinor and vector fields differs only by the numerical
coefficients $\al_{i}.$ Inserting appropriate coefficients listed in
the Table I into (7), one may easily generalize our discussion to the
fields of higher spins. Note however, that to obtain the appropriate
result for the neutral spinor field one has to multiply the
renormalized effective action by the factor 1/2.

\section{Electrically charged nonlinear black hole}

As is well known the Reissner-Nordstr\"om line element is the only
static and  asymptotically flat solution of the Einstein- Maxwell
equations representing a black hole of mass $M$ and electric charge
$e.$  The appropriate line element has the form
\begin{equation}
ds^{2}\,=\,-U(r) dt^{2}\,+\,V^{-1}(r)dr^{2}\,+\,
r^{2}\left( \sin^{2}\theta d\phi^{2}	\,+\,d\theta^{2}\right),
\end{equation}
where the metric functions $U(r)$ and $V(r)$ are given by
\begin{equation}
U(r)\,=\,V(r)\,=\,1\,-\,{2 M\over r}\,+\,{e^{2}\over r^{2}} .
\end{equation}
Because of its simplicity the RN solution may be studied
analytically; 
for $e^{2} < M^{2}$ the equation $g_{00}\,=\,0 $ has two positive roots
\begin{equation}
r_{\pm}\,=\,M\,\pm\,(M^{2}\,-\,e^{2})^{1/2},
\end{equation}
and the larger root represents the location of the event horizon,
while $r_{-}$ is the inner horizon. In the limit $e^{2}\,=\,M^{2}$
horizons merge at $r\,=\,M,$ and the RN solution
degenerates to the extremal one. 
The singularity of the RN line element that one encounters at $r=0$
is a non-removable curvature singularity, while those at $r_{\pm}$
are merely spurious singularities that may be easily removed
by a suitable choice of coordinates.

Recent interest in the nonlinear electrodynamics is partially
motivated, beside a natural curiosity,  by the fact that the theories
of this type frequently arise in modern theoretical physics. For
example they appear as effective theories of string/M-theory.
Moreover, one expects that it should be possible to construct
solutions to the coupled system of the Einstein field and equations of
the nonlinear electrodynamics, which may be interpreted as
representing globally regular black hole geometries, avoiding thus the
singularity problem. As the nonlinear electrodynamics in the weak
field limit coincides with  the Maxwell theory, one expects that the
appropriate solution should approach at large distances the
RN solution.

An interesting  solution of this type, representing spacetime of the
regular black hole with mass $M$ and charge $e$ has been constructed
recently by Ay\'on-Beato and Garc\'{\i}a~\cite{Beato}. The appropriate line
element is given by (25) with
\begin{equation}
U(r)\,=\,V(r)\,=\, 1 - {2 M\over r}\left[1 - \tanh \left({e^{2}\over 2 M r}
\right)\right].
\end{equation}
For $e=0$ the ABG solution reduces to the Schwarzschild solution;
 for small values of the charge it differs from the
Reisner-Nordstr\"om solution by  terms of order $O(e^{6}).$  At large
distances the metric structure of (28) also closely resembles that of
the RN solution. Indeed, expanding $U(r)$ in a
power series one concludes that the ABG solution behaves
asymptotically 	as
\begin{equation}
 U(r)\,=\,V(r)\,=\,1\,-\,{2 M\over r}\,+\,{e^{2}\over r^{2}}\,
 -\,{e^{6}\over 12 M^{2} r^{4}}\,+\,O({1\over r^{6}}).
 \end{equation}

Instead of referring to  numerical calculations at this stage of
analyses of the ABG geometry, we show that although the metric
coefficient $U(r)$ is a complicated function of $r$, the location of the
horizons may be elegantly expressed in terms of the Lambert
functions~\cite{Lambert}. Indeed, making use of the substitution $ r
= M x$ and $e^{2} = q^{2} M^{2},$ and subsequently introducing  a new
unknown function $W$ by means of the relation
\begin{equation}
x = - {4 q^{2}\over 4 W\,-\,q^{2}},
\end{equation}
one  arrives at
\begin{equation}
\exp (W) W = -{q^{2}\over 4}\exp ({q^{2}/4}) .
\end{equation}
 Since the Lambert function is defined
 as
 \begin{equation}
 \exp (W(s)) W(s)\,=\,s,
 \end{equation}
one concludes that the location of the horizons as a function of $q=|e|/M,$
is given by the real branches of the Lambert functions
\begin{equation}
x_{+}\,=\,- {4 q^{2}\over 4 W(0, -{q^{2}\over 4} \exp (q^{2}/4)) - q^{2}} ,
\end{equation}
and
\begin{equation}
x_{-}\,=\,- {4 q^{2}\over 4 W(-1, -{q^{2}\over4} \exp (q^{2}/4)) - q^{2}}.
\end{equation}
The functions $W(0,s)$ and $W(-1,s)$ are the only real branches of
the Lambert function with the branch point at $s = - 1/{\rm e},$
where e is the base of natural logarithms. The horizons $r_{+}$ and
$r_{-}$ for
\begin{equation}
|e|/M\,=\,2 W^{1/2}(1/{\rm e}),
\end{equation}
merge at 
\begin{equation}
x_{extr}\,=\,{4 W (1/{\rm e})\over 1 + W (1/{\rm e})},
\end{equation}
where $W(s)$ is a principal branch of the Lambert function $W(0,s).$
Numerically one has 
\begin{equation}
x_{extr}\,=\,0.871,
\end{equation}
and
\begin{equation}
{|e|\over M}\,=\,1.056.
\end{equation}
Inspection of (33) and (34) shows an interesting feature of the ABG
geometry: the black hole solution exists for $q$
greater than the analogous ratio of the parameters of the 
RN solution.

The location of $r_{+}$ and $r_{-}$ as a function of $q$ for the
charged black holes of both types are displayed in Fig. 1; its  
numerical values for some characteristic values of $|e|/M$ are presented
in Table II. Inspection of the figure shows that locations of the
event horizons of the RN and ABG solutions are almost indistinguishable
for, approximately, $ |e|/M \lesssim 0.7,$ whereas the differences
between the inner  horizons are more prominent. The latter differences
are irrelevant here as in our analyses we shall confine ourselves to the static
region exterior to the event horizon. Generally, for a given $q,$
$r_{+}$ of the RN black hole is always greater than $r_{+}$  of the
ABG geometry.

We remark here that the global structure of the ABG spacetime is
similar to that of RN, with one notable distinction. Simple analysis
shows that the curvature invariants of the curvature tensor, $R,$
$Ricci^{2},$ and $Riem^{2}$ are regular as $r\to 0,$ and, moreover,
other differential invariants of the Riemann tensor and its covariant
derivatives also exhibit regularity there. One concludes therefore
that the ABG geometry for $q \leq q_{extr}$ represents the regular
black hole solution. While this property of the ABG solution is not
surprising it should be remembered that earlier efforts have been in
unsuccessful in this regard.

By means of the Wick rotation one obtains the Euclidean version of
(25) with (28), which has no conical singularity provided the time
coordinate is periodic with the period given by
\begin{equation}
\beta_{H}\,=\,4\pi \lim_{r\to r_{+}}
\left[ U(r) V^{-1}(r)\right]^{1/2}\left({dU\over dr}\right)^{-1}.
 \end{equation}
Making use of elementary properties of the hyperbolic functions one has
\begin{equation}
\beta_{H}\,=\,4\pi 
\left[{1\over r_{+}}\,-\,{e^{2}\over r_{+}^{2} M}
\left( 1-{r_{+}\over 4 M}\right)\right]^{-1}.
 \end{equation}
We recall also that analogous period of the Euclideanized
 RN geometry is given by
 \begin{equation}
 \be_{H}\,=\,4\pi \left( {2 M\over r_{+}^{2}}\,-
 \,{2 e^{2}\over r_{+}^{3}}\right)^{-1}.
\end{equation}
In the limit $e \to 0$ both (40) and (41) tend to the Schwarzschild
value $8\pi M$ whereas in the extremality limit $\be_{H}$ tends to
infinity. As the Hawking temperature is proportional to the inverse of
the period $\be_{H}$ one concludes  that in  the extremality limit the
Hawking temperature of the ABG black hole vanishes. Moreover, closer
analysis indicates that for a given $e$ and $M$ the ABG black hole is
hotter than its RN counterpart characterized by the same values of the
parameters. Of course, as expected, for small electric charges both
temperatures are practically indistinguishable.

%%%%%%%%%%%%%%%%%%%%%%%%%%%%%%%%%
\section{ Renormalized stress-energy tensor in the spacetime of
electrically charged black hole}

 In this chapter the method	described in Sec. II is used to 
 construct the renormalized stress-energy tensor of the quantized
 massive scalar fields in the ABG and extremal ABG spacetimes
 in the region exterior to the event horizon.
As there are important simplifications for $\xi =1/6$
 we shall consider only the conformal coupling.

 The analogous tensor in the RN geometry has been
 evaluated in Ref.[8]  by means of the sixth-order WKB approximation of
 the solution to the scalar field equation and the summation thus
 obtained mode functions by means of the Abel-Plana formula.
 This result has been rederived and extended to the case of other
 spins, using simplified version of Eqs. (19-24) valid in the spacetimes
 with vanishing curvature scalar~\cite{kocio}.

 Calculating the components of the Riemann tensor, its contractions
 and required covariant derivatives, inserting the results into (9-18),
 performing appropriate simplifications, and finally constructing the
 renormalized stress-energy tensor, after rather lengthy calculations
 one has
\begin{eqnarray}
{1\over \gamma \alpha}\langle T^{t}_{t}\rangle_{ren}\, &=& \, - \, 
\Big\{ 576\,\gamma \,M^{7}\,r^{6}\,( - 626\,\gamma \,M + 285\,r)
   \nonumber \\
 & &  -\, 576\,\delta \,e^{2}\,M^{5}\,r^{5}\,(170\,\gamma ^{
2}\,M^{2} - 669\,\gamma \,M\,r + 270\,r^{2}) \nonumber \\
& &+ \,48\delta \,e^{4}\,
M^{4}\,r^{4} 
  \Big[ 186\,( - 47 + 181\,\beta )\,\gamma ^{2}\,M^{2} \nonumber \\
 & & +\, (3967 - 
38132\,\beta  + 34165\,\beta ^{2})\,M\,r 
+ 8520\,\beta \,r^{2}\Big] \nonumber \\
& & +\, 12\delta  
  e^{8}\,M^{2}\,r^{2}\Big[2 \,\gamma ^{2}\,M^{2} \,(  3431\,
\beta ^{2} + 47307\,\beta ^{3}- 2976 - 32264\,\beta  )\nonumber \\
 & &+ \,M\,r \,(2824 + 52712\,\beta  - 58927\,\beta ^{2} - 79068
\,\beta ^{3} + 82459\,\beta ^{4})\nonumber \\
 & &  +\, 5904\,\beta \,( - 2 + 3\,\beta ^{2})\,r^{2}\Big]  
\nonumber \\
& & - \,96\delta \,e^{6}\,M^{3}\,r^{3}\Big[6\,( - 1297 - 283\,\beta
  + 3576\,\beta ^{2})\,\gamma ^{2}\,M^{2} \nonumber \\
 & &  + \,(6730 - 5993\,\beta  - 20190\,\beta ^{2} + 19453\,
\beta ^{3})\,M\,r \nonumber \\
& &- \,1455\,r^{2}\,( 1 -3\,\beta ^{2})\Big]\mbox{}  \nonumber \\
& & -\, 12\delta \,e^{10}\,M 
  r\Big[2\,\gamma^{2}\,M^{2}(1302 - 2253\,\beta  - 12187\,\beta ^{2} \nonumber \\
  & & +\, 2485\,\beta 
^{3} + 12449\,\beta ^{4}) 
 \, -\,M\,r ( 2524 - 4653\,\beta  \nonumber \\
 & & -\, 18930\,\beta ^{2} + 
23420\,\beta ^{3} + 18930\,\beta ^{4} - 21291\,\beta ^{5})
 \nonumber \\
& &  + \, 298\,(2 - 15\,\beta ^{2} + 15\,\beta ^{4})\,r^{2}\Big]
\mbox{} \nonumber \\
& &+ \, \delta \,e^{12} \Big[
  2\,\gamma ^{2}\,M^{2}(693 + 4629\,\beta  - 5370\,\beta ^{2} - 19490\,\beta ^{3
} \nonumber \\
& &+ \,4745\,\beta ^{4} + 15409\,\beta ^{5}) 
  - 3\,M\,r ( 231 + 2720\,\beta  - 4645\,\beta ^{2}\nonumber \\
 & & -\, 
9600\,\beta ^{3} + 12785\,\beta ^{4} + 7200\,\beta ^{5} - 8691\,
\beta ^{6}) \nonumber \\
 & &  + \,120\,\beta \,r^{2}(17 - 60\,\beta ^{2} + 45\,\beta ^{4}
)\Big]	 \Big\},
\end{eqnarray}
%%%%%%%%%%%%%%%%%%%%
%%%%%%%%%%%%%%%%%%%%
%%%%
\begin{eqnarray}
{1\over \gamma M \alpha}\lefteqn{\langle T^{r}_{r}\rangle_{ren}\,=\,4032\,\gamma \,M^{6}\,(22
\,\gamma \,M - 15\,r)\,r^{6} }\nonumber \\
& &
+ \delta ^{2}\,e^{12}\,\gamma \Big[
 (270 - 558\,\beta  - 1500\,\beta ^{2} + 2060\,\beta ^{3} + 
1350\,\beta ^{4} - 1622\,\beta ^{5})\,M   \nonumber \\
& &  - 15\,(9 - 50\,\beta ^{2} + 45\,\beta ^{4})\,r\Big]\nonumber \\
& &
+ 12\delta \,e^{2}\,r \Big\{
  - 48\,M^{4}\,r^{4}\,(478\,\gamma ^{2}\,M^{2} - 391\,\gamma 
\,M\,r + 54\,r^{2}) \nonumber \\
& &
+ 4e^{2}\,M^{3}\,r^{3} 
  \Big[ 6\,(101 + 873\,\beta )\,\gamma ^{2}\,M^{2} 
  \nonumber \\
& &  -( 707 + 
3148\,\beta  - 3855\,\beta ^{2})\,M r + 472\,\beta \,r^{2}  \Big]\mbox{}\nonumber \\
& &
+ e^{6} M\,r \Big[ 2\,( - 544 - 1304\,\beta  + 957\,\beta ^{2} + 2041\,
\beta ^{3})\,\gamma ^{2}\,M^{2} \nonumber \\
 & & + r\,(392 + 408\,\beta  - 1717\,\beta ^{2} - 612\,
\beta ^{3} + 1529\,\beta ^{4})\,M + 40\,\beta \,r\,(2 - 3\,\beta ^{2
})\Big]\mbox{}  \nonumber \\
 & & - 8e^{4}\,M^{2}\,r^{2} \Big[ 6\,( - 73 + 35\,\beta  + 222\,\beta ^{2
})\,\gamma ^{2}\,M^{2} \nonumber \\
 & & + r\,(170 - 387\,\beta  - 510\,\beta ^{2} + 727\,
\beta ^{3})\,M + 11\,(  3\,\beta ^{2} - 1)\,r^2 \Big]\mbox{}\nonumber \\
& &
+ e^{8} \Big[
  2\,( 18 + 407\,\beta  + 441\,\beta ^{2} - 487\,\beta ^{
3} - 551\,\beta ^{4})\,\gamma ^{2}\,M^{2}\nonumber \\
& & + r 
 (4 - 287\,\beta  - 30\,\beta ^{2} + 676\,\beta ^{3} + 30\,
\beta ^{4} - 393\,\beta ^{5})\,M  \nonumber \\
& &
+ 2\,(2 - 15\,\beta ^{2} + 15\,
\beta ^{4})\,r^{2}  
 \Big] \Big\},
\end{eqnarray}
%%%%%%%%%%%%%%%%%%%%%
and
%%%%%%%%%%%%%%%%%%%%%
 \begin{eqnarray}
{1\over \ga \al}\lefteqn{\langle T^{\theta}_{\theta}\rangle_{ren} = 576\,\gamma \,M^{7}\,
(
734\,\gamma \,M - 315\,r
)
\,r^{6} }\nonumber \\
& &
+ \delta \,e^{2} 
\Big\{ 
  - 576\,M^{5}\,r^{5}\,(2202\,\gamma ^{2}\,M^{2} - 1313\,
\gamma \,M\,r + 162\,r^{2}
) \nonumber \\
& &
+ 48e^{2}\,M^{4}\,r^{4}
\Big[ 
  6\,(1399 + 3777\,\beta )\,\gamma ^{2}\,M^{2} + r\,( - 3353
 - 10914\,\beta  + 14267\,\beta ^{2})\,M 
 \nonumber \\
 & & + 1976\,\beta \,r^{2} \Big]
  - 12e^{6}\,M^{2}\,r^{2} \Big[ 2\,( - 202 - 1198\,\beta 
 + 749\,\beta ^{2} + 2161\,\beta ^{3})\,\gamma ^{2}\,M^{2} 
 \nonumber \\
 & &
 +  r\,(154 + 1184\,\beta  - 2227\,\beta ^{2} - 1776\,\beta ^{3
} + 2665\,\beta ^{4})\,M + 92\,\beta \,( - 2 + 3\,\beta ^{2})\,r^2
\Big] \nonumber \\
 & & - 48e^{4}\,M^{3}\,r^{3}\Big[ 6\,( - 91 + 649\,\beta  + 
1042\,\beta ^{2})\,\gamma ^{2}\,M^{2} \nonumber \\
 & & \mbox{} + r\,(873 - 2308\,\beta  - 2619\,\beta ^{2} + 4054
\,\beta ^{3})\,M + 206\,( - 1 + 3\,\beta ^{2})\,r^{2}\Big]\mbox{} \nonumber \\
& &
+ 12e
^{8} M\,r \Big[ 2\,(80 - 519\,\beta  - 1169\,\beta ^{2} + 1089\,\beta 
^{3} + 1755\,\beta ^{4})\,\gamma ^{2}\,M^{2} \nonumber \\
& &
+ r ( - 176 + 639\,\beta  + 1320\,\beta ^{2} - 2716\,\beta ^{3}
 - 1320\,\beta ^{4} + 2253\,\beta ^{5})\,M \nonumber \\
 & &  + 20\,(2 - 15\,\beta ^{2} + 15\,\beta ^{4})\,r^{2}\Big]\nonumber \\
& & - e^{10}\Big[ 
  2\,(129 + 441\,\beta  - 1590\,\beta ^{2} - 2750\,\beta ^{3}
 + 2105\,\beta ^{4} + 3001\,\beta ^{5})\,\gamma ^{2}\,M^{2} \nonumber \\
 & &
 + 3r ( - 43 - 272\,\beta  + 781\,\beta ^{2} + 960\,\beta ^{3} - 
2005\,\beta ^{4} - 720\,\beta ^{5} + 1299\,\beta ^{6})\,M \nonumber \\
 & & \mbox{} + 4\,\beta \,(17 - 60\,\beta ^{2} + 45\,\beta ^{4})
\,r^{2}\Big]
\Big\},
\end{eqnarray}
where
\begin{equation}
\alpha^{-1}\,=\,96\pi^{2}m^{2}\times 60480 M^{5} r^{15},
\end{equation}
\begin{equation}
\beta\,=\,\tanh {e^{2}\over 2 M r},
\end{equation}
\begin{equation}
\gamma\,=\,1-\beta,
\end{equation}
and
\begin{equation}
\delta\,=\,1+\beta .
\end{equation}
%%%%%%%%%%%%%%%%%%%%%%%%  %%%%%%%%%%%%%%%%%
%%%%%%%%%%%%%%%%%%%%%%%%%%%%%%%%%%%%%%%%%%%
Obtained tensor is, as expected,  covariantly conserved, and as could
be easily verified  in the limit $e = 0,$ it reduces to the
stress-energy tensor constructed in the Schwarzschild spacetime by
Frolov and Zel'nikov.

Equations (8-18) may be employed also in the RN geometry. Since the
scalar curvature is zero there,  both $T^{(2)ab}$ and $T^{(3)ab}$
vanishes, and the resulting tensor exhibits simple linear dependence
on the coupling constant. Indeed, repeating the calculations for the
line element (26), one has~[8,16]
\begin{equation}
  \langle T^{b}_{a}\rangle_{ren}^{(0)}\,=\,C^{b}_{a}\,+
  \,\left(\xi	\,-\,{1\over 6}\right) D^{b}_{a},
\end{equation}
where
 \begin{eqnarray}
 C^{t}_{t}\,&=&\,-{\frac {1}{30240\,\pi^{2}\,m^{2}\, r^{12}}}\,\left( 1248\,{e}^{6}-810\,
 {r}^{4}{e}^{2}+855\,{M}^{2}{r}^{4}+202\,{r}^{2}{e}^{4}  \right. \nonumber \\
&-&\left. \,1878\,{M}^{3}{r}^{3}+1152\,M{r}^{3}{e
}^{2}+2307\,{M}^{2}{r}^{2}{e}^{2}-3084\,r M{e}^{4}\right),
 \end{eqnarray}

\begin{eqnarray}
C^{r}_{r}\,&=&\,\frac {1}{30240\, \pi^{2}\,m^{2}\,r^{12}}\left( 444\,{e}^{6}
-1488\,M{r}^{3}{e}^{2}+162\,{r
}^{4}{e}^{2}
+\,842\,{r}^{2}{e}^{4}-1932\,r M{e}^{4}\right. \nonumber \\
&+&\left.\,315\,{M}^{2}{r}^{4}+
2127\,{M}^{2}{r}^{2}{e}^{2}-462\,{M}^{3}{r}^{3}\right),
\end{eqnarray}
 and
\begin{eqnarray}
 C^{\theta}_{\theta}\,&=&\,-\,\frac {1}{30240\,\pi^{2}\,m^{2}\,r^{12}}
 \,\left( 3044\,{r}^{2}{e}^{4}-2202\,{M}^{3}{r}^{3}
-10356\,r M {e}^{4} \right. \nonumber \\
&+&\left.\,
3066\,{e}^{6}-4884\,{r}^{3}M{e}^{2}+9909\,{r}^{2}{M}
^{2}{e}^{2}+945\,{M}^{2}{r}^{4}+486\,{r}^{4}{e}^{2}\right).
\end{eqnarray}
Since we are interested in the conformally coupled massive scalar
fields, the exact form of the $D_{a}^{b}$ tensor is irrelevant.

Now, we shall address the  question of how the differences between the
geometry of the black hole spacetimes constructed within the framework
of the Einstein-Maxwell theory on the one hand  and the nonlinear
electrodynamics coupled to the General Relativity  on the other, are
reflected in the overall behavior of the components of the
stress-energy tensors. To answer this, let us analyze numerically
$\langle T_{a}^{b}\rangle_{ren}$ in both cases. The results of our
calculations are presented graphically in Figs. 2 - 10. The plots of
the  time, radial, and angular components of the stress-energy tensor
of the quantized massive scalar field as a functions of the (rescaled)
radial coordinate in the spacetimes of ABG and RN black holes for
three exemplar values of the ratio $|e|/M\,=\,0.1,\, 0.5\,{\rm and}
\,0.95,$ are supplemented by similar plots drawn for the extremal
black holes. Inspection of the figures indicates that there are
striking qualitative similarities between the RN and ABG solutions for
a given $q$. Moreover, for  small values of the ratio the curves are
practically undistinguishable from each other, and, as expected,
noteworthy differences occur only for the black holes at and  near the
extremality limit. Since at large distances the line element (28)
approaches that of the RN, the most interesting region is the
neighborhood of the black hole event horizon. From Eq.(19) we know that
the renormalized stress-energy tensor depends on the coupling constant
$\xi$  in a  complicated way, and, therefore, one should not
expect that such similarities occur also in a general case.

Specifically, the dependence  of $\langle
T_{t}^{t}\rangle_{ren}$  constructed in the spacetime of the nonlinear
black hole and the RN geometry on $r/r_{+}$  for $|e|/M = 0.1, 0.5,
0.95$, are shown in Figs. 2 and 3, respectively. In Fig. 4  similar
curves are drawn for the extremal black holes. In the most interesting
region, i.e., in the vicinity of the event horizon, the energy
density, $\rho\,=\,-\langle T_{t}^{t}\rangle_{ren},$ is negative and
decreases with increasing of $|e|/M.$

As is seen in Figs. 5 -7, the radial component of the stress-energy
tensor is everywhere positive, and the horizon values of the radial
pressure, $p_{r}\,=\,\langle T_{r}^{r}\rangle_{ren},$ increases with
increasing $|e|/M.$

Of all components of the renormalized stress-energy tensor, the most
complicated behavior exhibits the angular pressure
$p_{\theta}\,=\,\langle T^{\theta}_{\theta}\rangle_{ren}$ (Figs 8 - 10). Indeed, for
the ABG black hole the angular pressure is positive on the event
horizon for $|e|/M \lesssim \,0.937$ and negative for larger values of
the ratio. Moreover, for $q \,\lesssim\, 0.903,$ $\langle
T^{\theta}_{\theta}\rangle_{ren}$ has a maximum at $r = r_{+},$
whereas for larger values the angular pressure has its maximum away
from the event horizon. Similarly, for the RN black hole $p_{\theta}$
is positive for $q\,\lesssim\,0.927$ and it has its maximum away from
the event horizon for $q\,\gtrsim\,0.864.$

It could be checked by a direct calculation that
\begin{equation}
\lim_{r\to r_{+}}\left(\langle T_{t}^{t}\rangle_{ren} \,-
\,\langle T_{r}^{r}\rangle_{ren}\right)
\left\{ 1- {2 M\over r}\left[1-\tanh {e^2\over 2 M r})\right]
\right\}^{-1}
\end{equation}
remains finite at the event horizon.
We observe that since the DeWitt-Schwinger approximation is
local and the geometry at the event horizon is regular, one expects
that that the renormalized stress-energy tensor is also regular there.

It should be stressed once again that for arbitrary curvature coupling
one has to incorporate also the terms $T^{(1)ab},$ $T^{(2)ab},$  and
$T^{(3)ab},$ that may considerably modify the  results. Moreover,
inspection of the Table I shows that for the neutral massive spinor
and vector fields in the ABG spacetime one has to use the  full system
(9 - 18) while in the geometry of the RN black hole, the terms (9) and
(11) do not contribute to the final result.

\section{Concluding remarks}

 In this paper we have constructed the renormalized stress-energy
 tensor of the massive conformally coupled scalar fields in the
 spacetime of the electrically charged black hole, being the solution
 of the coupled Einstein equation and the equation of nonlinear
 electrodynamics. A regular solution of  this type has been recently
 given by Ay\'on-Beato and Garc\'{\i}a. The method employed here is based on
 the observation that the first order effective action  could be
 expressed in terms of the traced coincidence limit of the coefficient
 $a_{3}.$ The general $ \langle T^{b}_{a}\rangle_{ren},$
 which has been obtained by functional differentiation of the
 effective action with respect to a metric tensor, has been applied in
 the spacetime of the nonlinear black hole. Since the
 Reissner-Nordstr\"om line element and ABG solution are practically
 indistinguishable from  each other for small values of $|e|/M,$ one
 expects that this similarity should be reflected in the behavior of
 the renormalized stress-energy tensor. Explicit calculations confirm
 this hypothesis and show that important differences between
 appropriate tensors, $\langle T_{a}^{b}\rangle_{ren},$ evaluated in
 the spacetime of the RN  black hole and that of ABG occur, as
 expected, near the extremality limit. For small $q$ constructed
 tensors are practically indistinguishable. Moreover, analyses of the
 Hawking temperatures indicate that for a given mass and electric
 charge, the ABG black hole is hotter than its RN black hole
 counterpart. Since  notable differences appear for temperatures close
 to zero one can ascribe this to the different ways of approaching the
 extremality limits.

 Apart from obvious extensions of our results to the massive scalar
 fields with arbitrary curvature coupling and to fields of higher
 spins, let us mention an interesting and important direction for
 future work. It is a problem of the back reaction of the quantized
 fields upon spacetime geometry of the ABG black hole, which may be
 studied perturbatively by means of the semiclassical Einstein field
 equations with a source term given by the renormalized stress-energy
 tensor of the quantized massive field and the classical stress-energy
 tensor of the background nonlinear electromagnetic field. To
 guarantee the renormalizabilty at that level, the semiclassical
 equations should contain higher derivative geometric terms. It is
 especially important in view of the recent  claim that the
 semiclassical zero temperature RN black holes do not exists [14].
 
 It should be stressed that the DeWitt-Schwinger expansion is local,
 and, therefore, does not describe particle creation which is a
 nonperturbative and nonlocal phenomenon. The method also breaks down
 in strong or rapidly varying gravitational fields, and, moreover, the
 massless limit leads to the nonphysical divergences. However, it is
 expected that for sufficiently massive scalar field the
 DeWitt-Schwinger approximation provides a good approximation of the
 exact renormalized stress-energy tensor.

%%%%%%%%%%%%
%%%%%%%%%%%% FIGURES
%%%%%%%%%%%%
 %%%%%%%%% Figure 1 %%%%%%%%%%%%%%
\begin{figure}
\centerline{
\framebox{
\psfig{figure=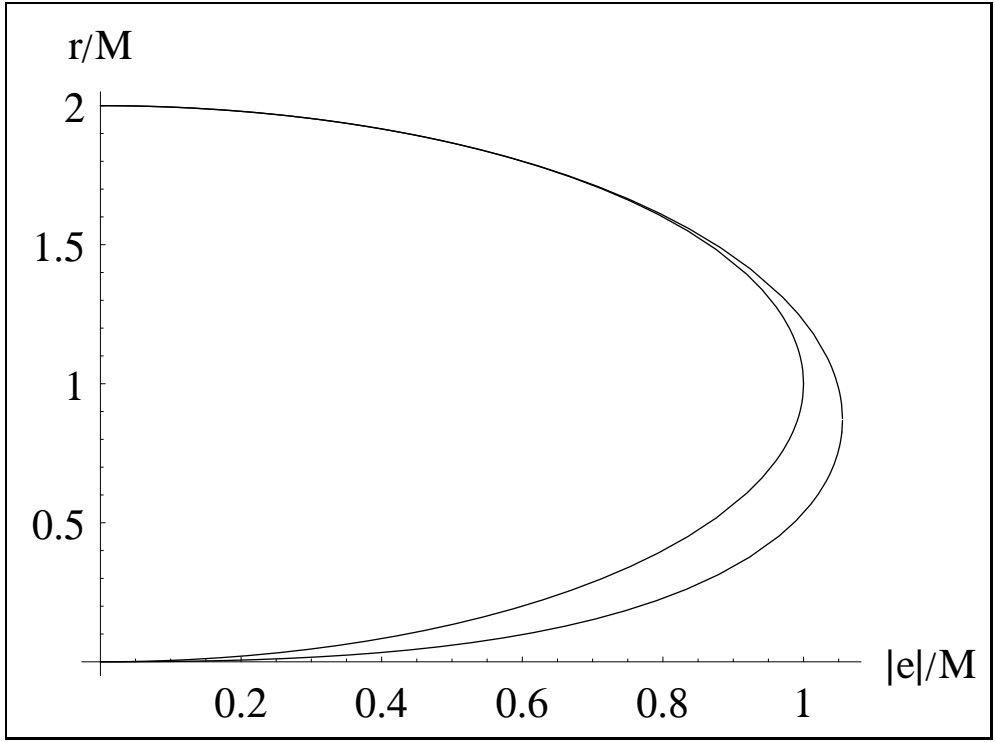}}}%,height=5cm,width=8cm}}}
\vspace{5mm}
\caption{The location of $r_{+}$ (upper branches) and $r_{-}$ (lower branches) of
the RN and ABG geometries as a function of $e/M.$
The curve representing horizons of the ABG black hole is shifted to the 
right with respect to the one wich has been determined in the RN
spacetime.
} 
\label{fig1}
\end{figure}
\clearpage
%%%%%%%%%%%%%%%%%%%%%%%%%%%%%%%
 %%%%%%%%%%% figures 2 - 10 %%%%%%%%%%%%%%%
 %%%%%%%%% Figure 2 %%%%%%%%%%%%%%
 \begin{figure}
 \centerline{
 \framebox{
 \psfig{figure=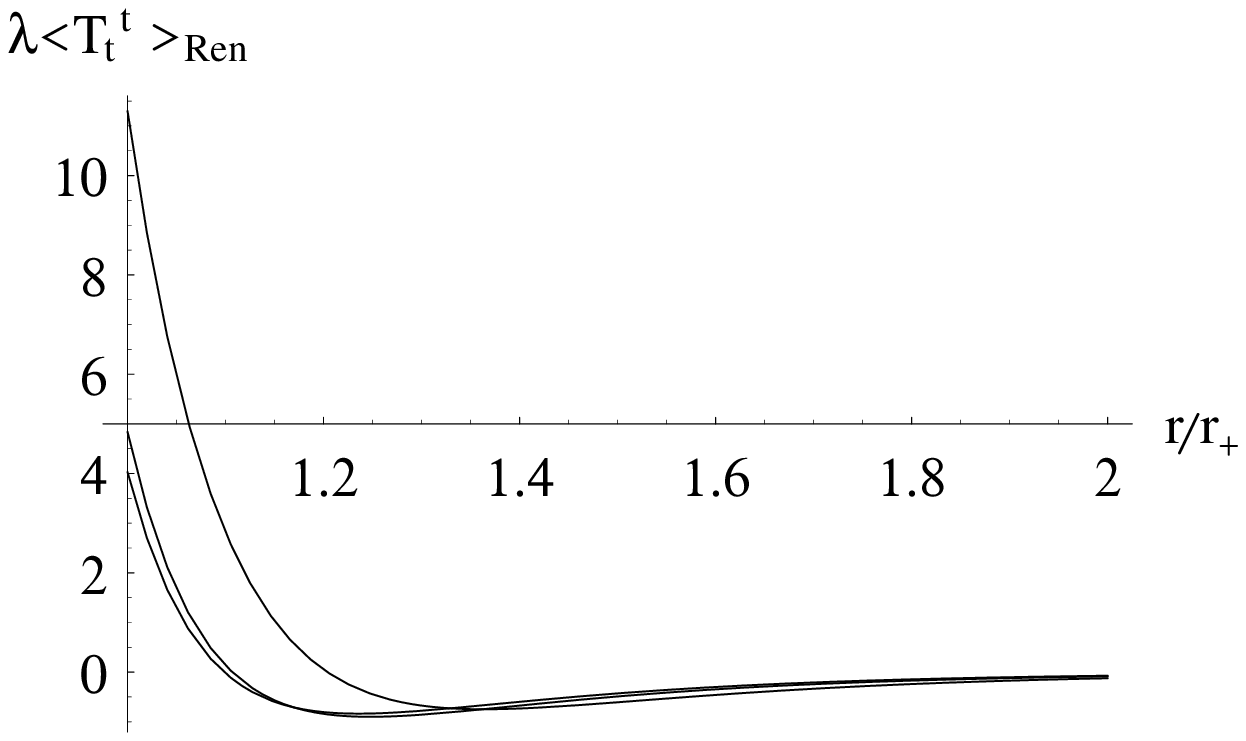}}}%,height=5cm,width=8cm}}}
 \vspace{5mm}
 \caption{This graph shows the radial dependence of the rescaled
 component $\langle T_{t}^{t}\rangle_{ren}$ $[\la = 90 (8 M)^{4}
 m^{2}\pi^{2}]$ of the renormalized stress-energy tensor of the
 massive conformally coupled scalar field in the ABG
 geometry. From top to bottom the curves are for $|e|/M =
 0.95,\,0.5,\,{\rm and}\,0.1.$ In each case $\langle
 T_{t}^{t}\rangle_{ren}$ has its positive maximum at $r = r_{+}$ and
 attains negative minimum away from the event horizon. }
 \label{fig2}
 \end{figure}
 %%%%%%%%%%%%%%%%%%%%%%%%%%%%%%%
 %%%%%%%%% Figure 3 %%%%%%%%%%%%%%
 \begin{figure}
 \centerline{
 \framebox{
 \psfig{figure=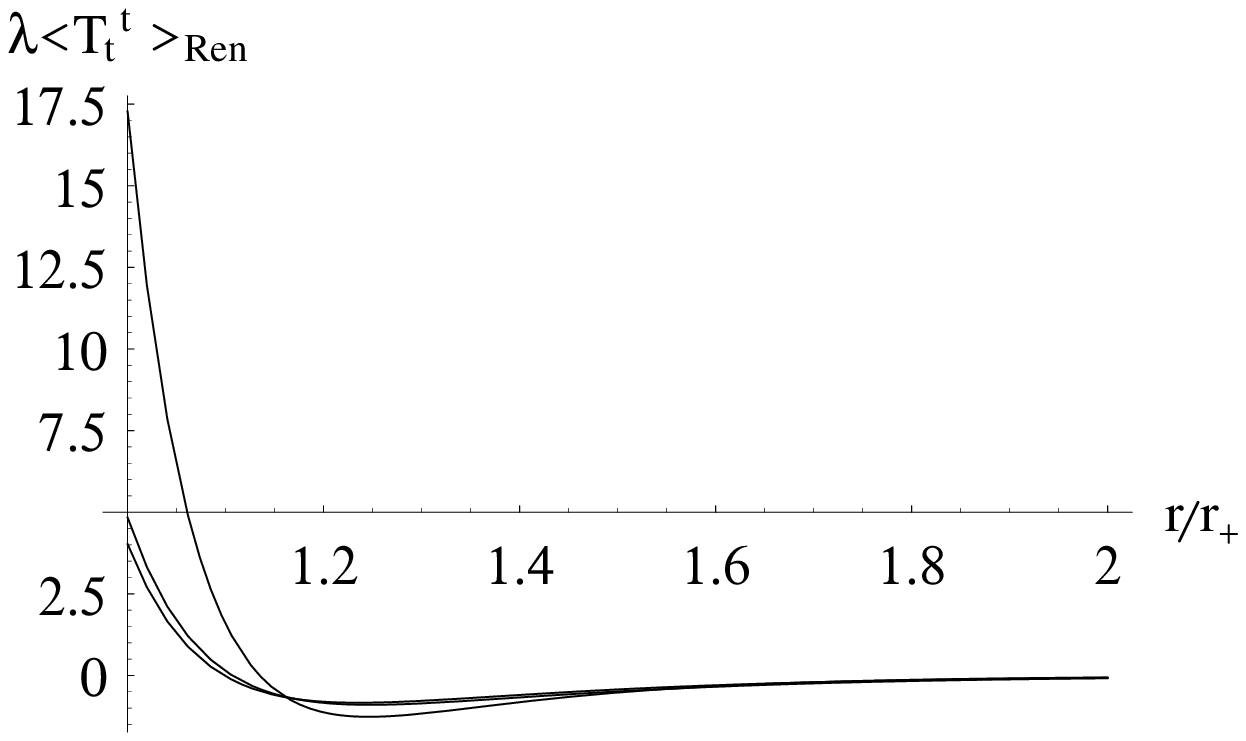}}}%,height=5cm,width=8cm}}}
 \vspace{5mm}
 \caption{This graph shows the radial dependence of the rescaled
 component $\langle T_{t}^{t}\rangle_{ren}$ $[\la = 90 (8 M)^{4}
 m^{2}\pi^{2}]$ of the renormalized stress-energy tensor of the
 massive conformally coupled scalar field in the RN
 spacetime. From top to bottom the curves are for $|e|/M =
 0.95,\,0.5,\,{\rm and}\,0.1.$ In each case $\langle
 T_{t}^{t}\rangle_{ren}$ has its positive maximum at $r = r_{+}$ and
 attains negative minimun away from the event horizon.} 
 \label{fig3}
 \end{figure}
 %%%%%%%%%%%%%%%%%%%%%%%%%%%%%%%
  %%%%%%%%% Figure 4 %%%%%%%%%%%%%%
  \begin{figure}
  \centerline{
  \framebox{
  \psfig{figure=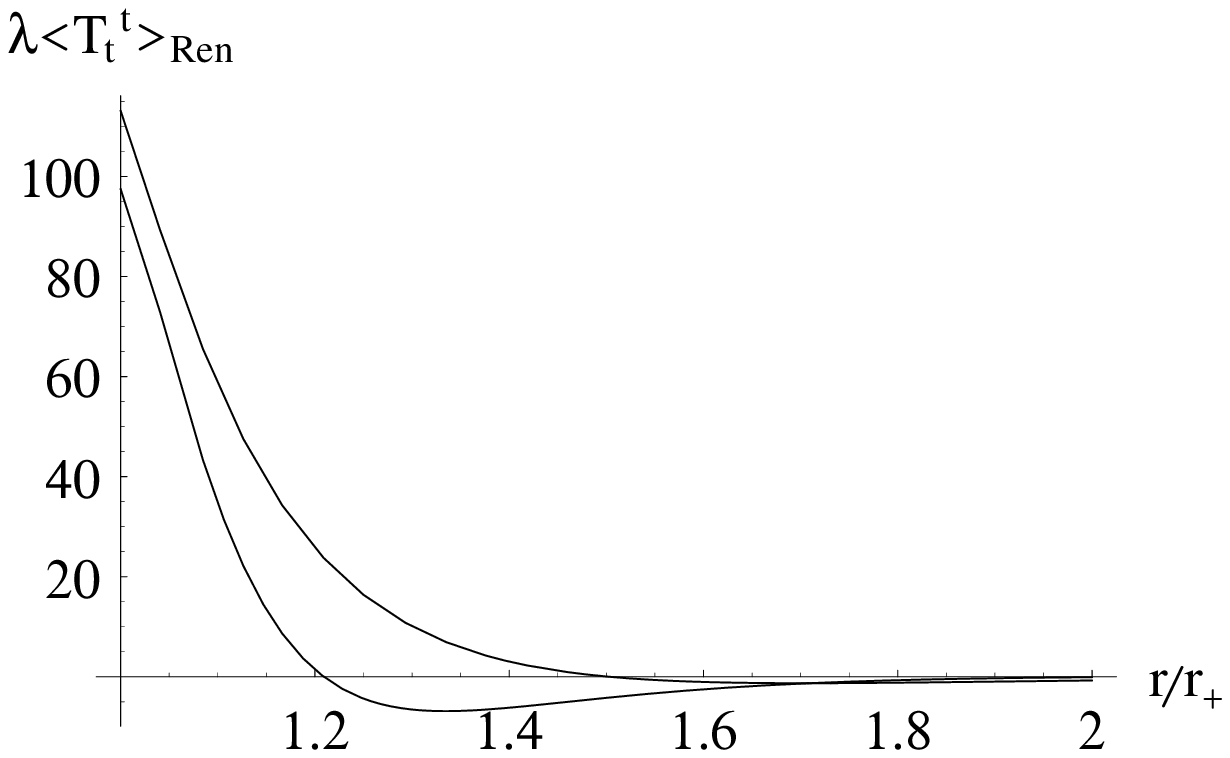}}}%,height=5cm,width=8cm}}}
  \vspace{5mm}
  \caption{This graph shows the radial dependence of the rescaled
 component $\langle T_{t}^{t}\rangle_{ren}$ $[\la = 90 (8 M)^{4}
 m^{2}\pi^{2}]$ of the renormalized stress-energy tensor of the
 massive conformally coupled scalar field. Top to bottom the curves are 
 respectively for the extremal ABG geometry and the 
 extremal RN black hole} 
  \label{fig4}
  \end{figure}
  %%%%%%%%%%%%%%%%%%%%%%%%%%%%%%%
  %%%%%%%%% Figure 5 %%%%%%%%%%%%%%
  \begin{figure}
  \centerline{
  \framebox{
  \psfig{figure=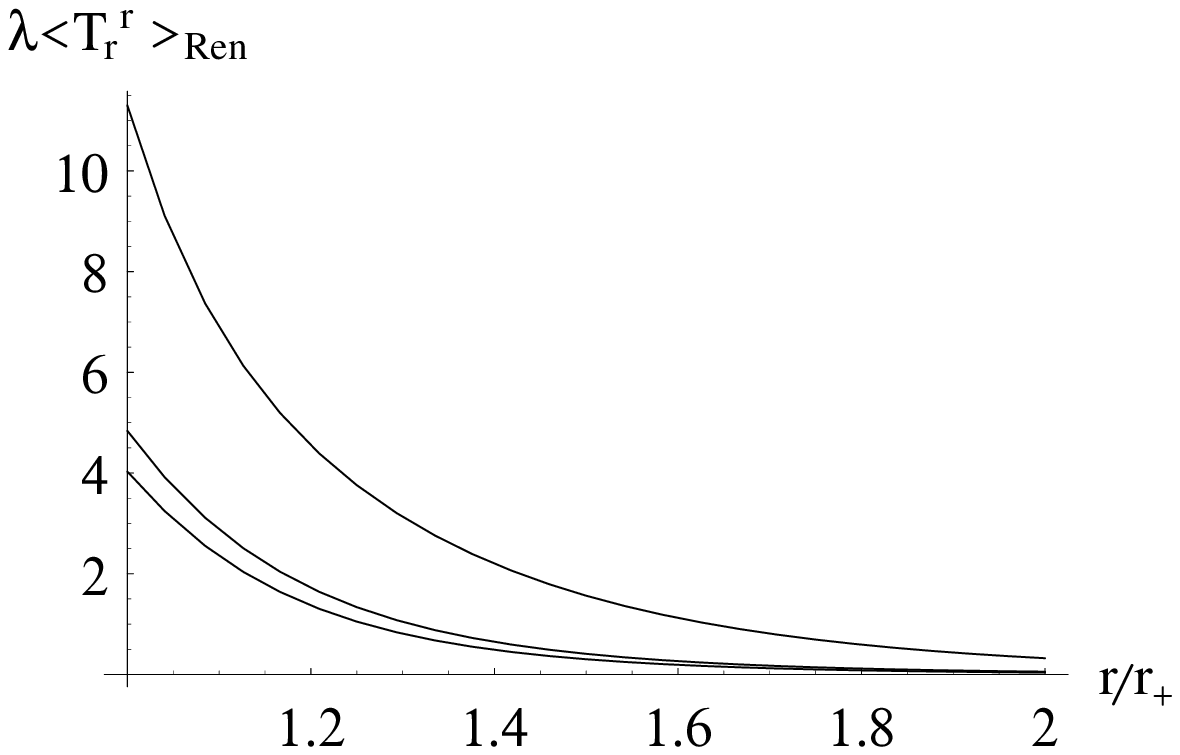}}}%,height=5cm,width=8cm}}}
  \vspace{5mm}
  \caption{This graph shows the radial dependence of the rescaled
 component $\langle T_{r}^{r}\rangle_{ren}$ $[\la = 90 (8 M)^{4}
 m^{2}\pi^{2}]$ of the renormalized stress-energy tensor of the
 massive conformally coupled scalar field in the Ay\'on-Beato and Gracia
 geometry. From top to bottom the curves are for $|e|/M =
 0.95,\,0.5,\,{\rm and}\,0.1.$ In each case $\langle
 T_{r}^{r}\rangle_{ren}$ has its positive maximum at $r = r_{+}$ and 
 decreases monotonically with $r$.} 
  \label{fig5}
  \end{figure}
  %%%%%%%%%%%%%%%%%%%%%%%%%%%%%%%
  %%%%%%%%% Figure 6 %%%%%%%%%%%%%%
  \begin{figure}
  \centerline{
  \framebox{
  \psfig{figure=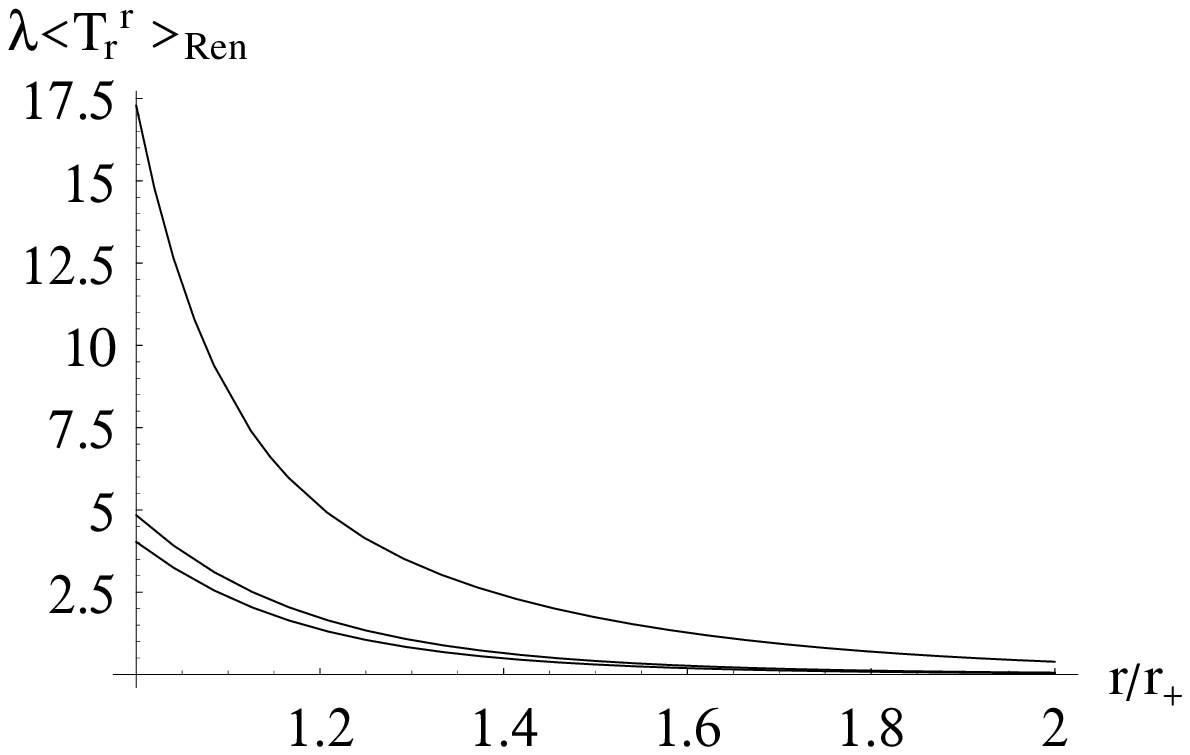}}}%,height=5cm,width=8cm}}}
  \vspace{5mm}
  \caption{This graph shows the radial dependence of the rescaled
 component $\langle T_{r}^{r}\rangle_{ren}$ $[\la = 90 (8 M)^{4}
 m^{2}\pi^{2}]$ of the renormalized stress-energy tensor of the
 massive conformally coupled scalar field in the RN
 spacetime. From top to bottom the curves are for $|e|/M =
 0.95,\,0.5,\,{\rm and}\,0.1.$ In each case $\langle
 T_{t}^{t}\rangle_{ren}$ has its positive maximum at $r = r_{+}$ and
 decreases monotonically with $r$.} 
  \label{fig6}
  \end{figure}
  %%%%%%%%%%%%%%%%%%%%%%%%%%%%%%%
  %%%%%%%%% Figure 7 %%%%%%%%%%%%%%
  \begin{figure}
  \centerline{
  \framebox{
  \psfig{figure=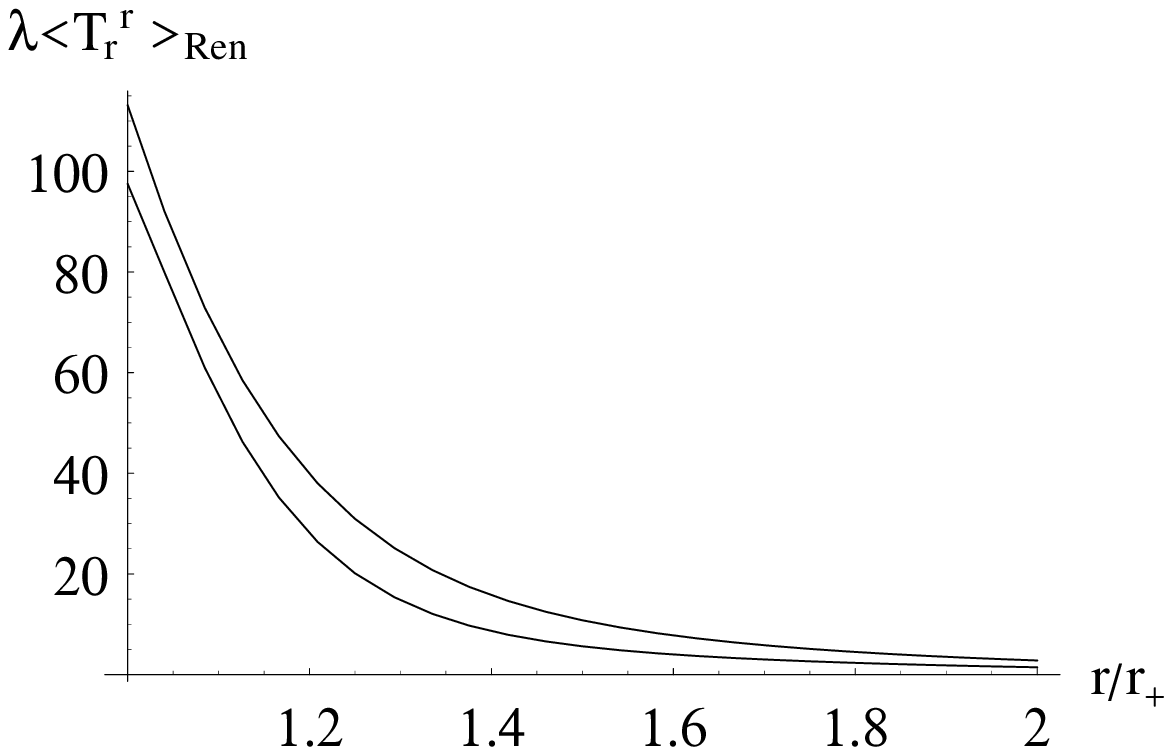}}}
  \vspace{5mm}
  \caption{This graph shows the radial dependence of the rescaled
 component $\langle T_{r}^{r}\rangle_{ren}$ $[\la = 90 (8 M)^{4}
 m^{2}\pi^{2}]$ of the renormalized stress-energy tensor of the
 massive conformally coupled scalar field. Top to bottom the curves are 
 respectively for the extremal  ABG geometry and the 
 extremal RN black hole} 
  \label{fig7}
  \end{figure}
  %%%%%%%%%%%%%%%%%%%%%%%%%%%%%%%
  %%%%%%%%% Figure 8 %%%%%%%%%%%%%%
  \begin{figure}
  \centerline{
  \framebox{
  \psfig{figure=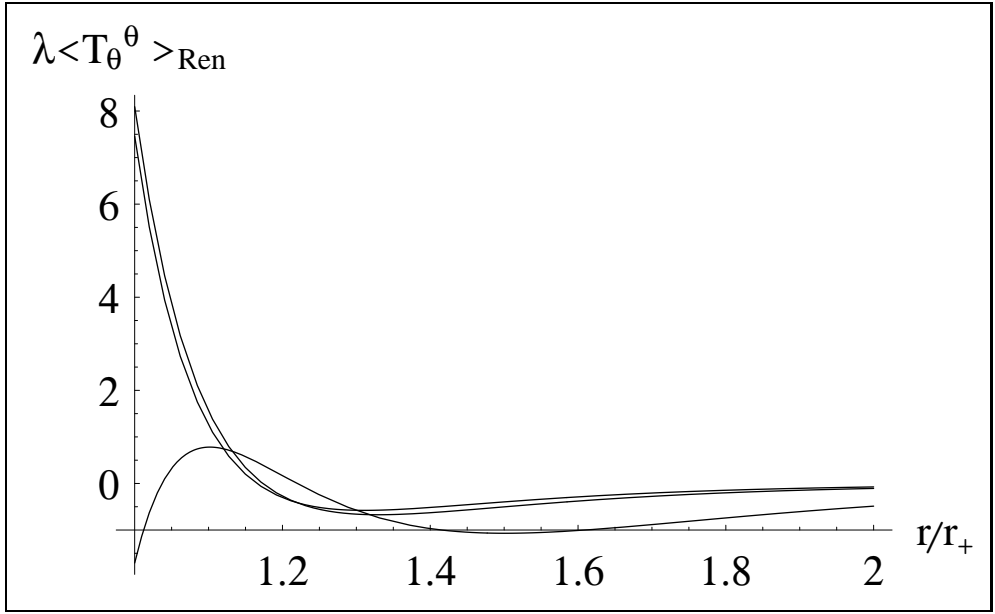}}}
  \vspace{5mm}
  \caption{This graph shows the radial dependence of the rescaled component  
  $\langle T_{\theta}^{\theta}\rangle_{ren}$ 
 $[\la = 90 (8M)^{4} m^{2}\pi^{2}]$ of the renormalized stress-energy tensor of the
 massive conformally coupled scalar field in the ABG
 geometry. From top to bottom the curves are for $|e|/M =
 0.1,\,0.5,\,{\rm and}\,0.95.$ For $|e|/M\,\leq\,0.903,$ $\langle
 T_{\theta}^{\theta}\rangle_{ren}$ has its positive maximum at $r =
 r_{+}.$ For larger values of the ratio it approaches its maximum away
 from the event horizon.  For $|e|/M = 0.937$ the angular pressure vanishes
 on the event horizon.} 	
  \label{fig8}
  \end{figure}
  %%%%%%%%%%%%%%%%%%%%%%%%%%%%%%%
  %%%%%%%%% Figure 9 %%%%%%%%%%%%%%
  \begin{figure}
  \centerline{
  \framebox{
  \psfig{figure=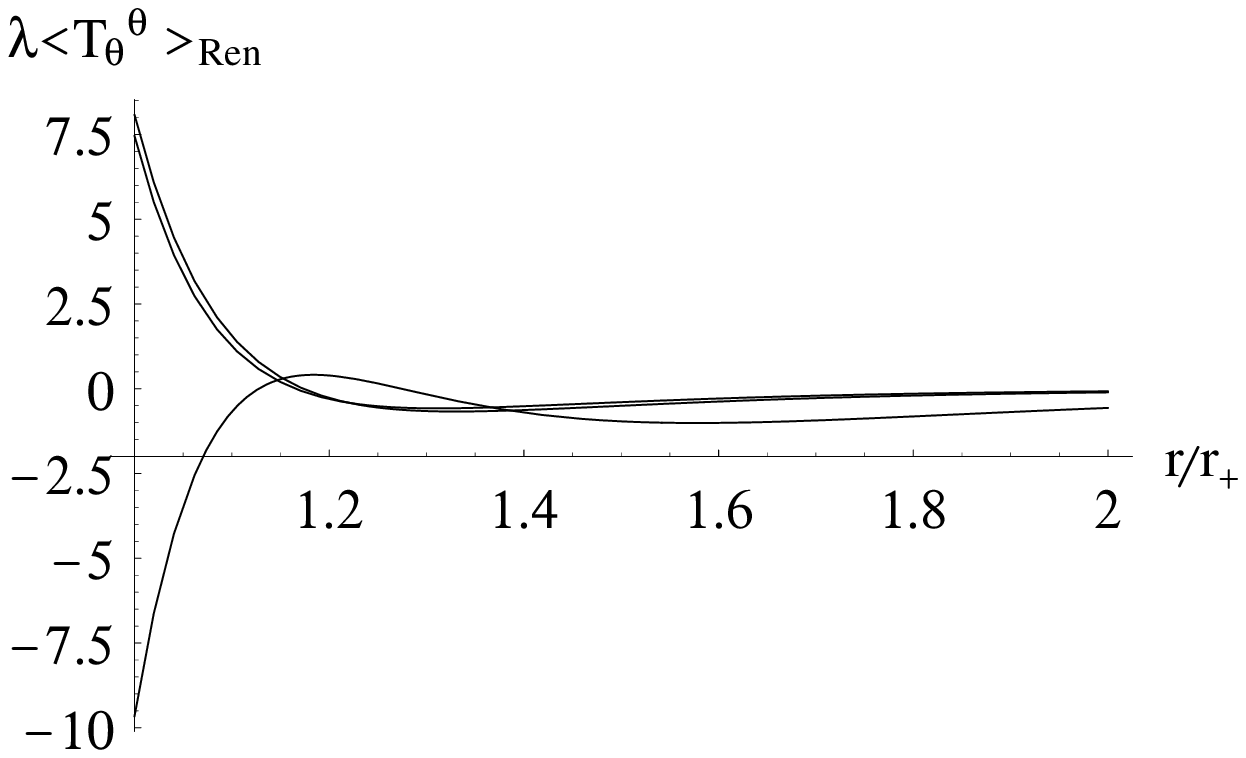}}}
  \vspace{5mm}
  \caption{This graph shows the radial dependence of the rescaled component 
  $\langle T_{\theta}^{\theta}\rangle_{ren}$ $[\la = 90 (8
 M)^{4} m^{2}\pi^{2}]$ of the renormalized stress-energy tensor of the
 massive conformally coupled scalar field in the RN
 spacetime. From top to bottom the curves are for $|e|/M =
 0.1,\,0.5,\,{\rm and}\,0.95.$ For $|e|/M\,\leq\,0.864,$ $\langle
 T_{\theta}^{\theta}\rangle_{ren}$ has its positive maximum at $r =
 r_{+}.$ For larger values of the ratio it approaches its maximum away
 from the event horizon.  For $|e|/M = 0.927$ the angular pressure vanishes
 at $r_{+}.$} 
  \label{fig9}
  \end{figure}
  %%%%%%%%%%%%%%%%%%%%%%%%%%%%%%%
   %%%%%%%%% Figure 10 %%%%%%%%%%%%%%
   \begin{figure}
   \centerline{
   \framebox{
   \psfig{figure=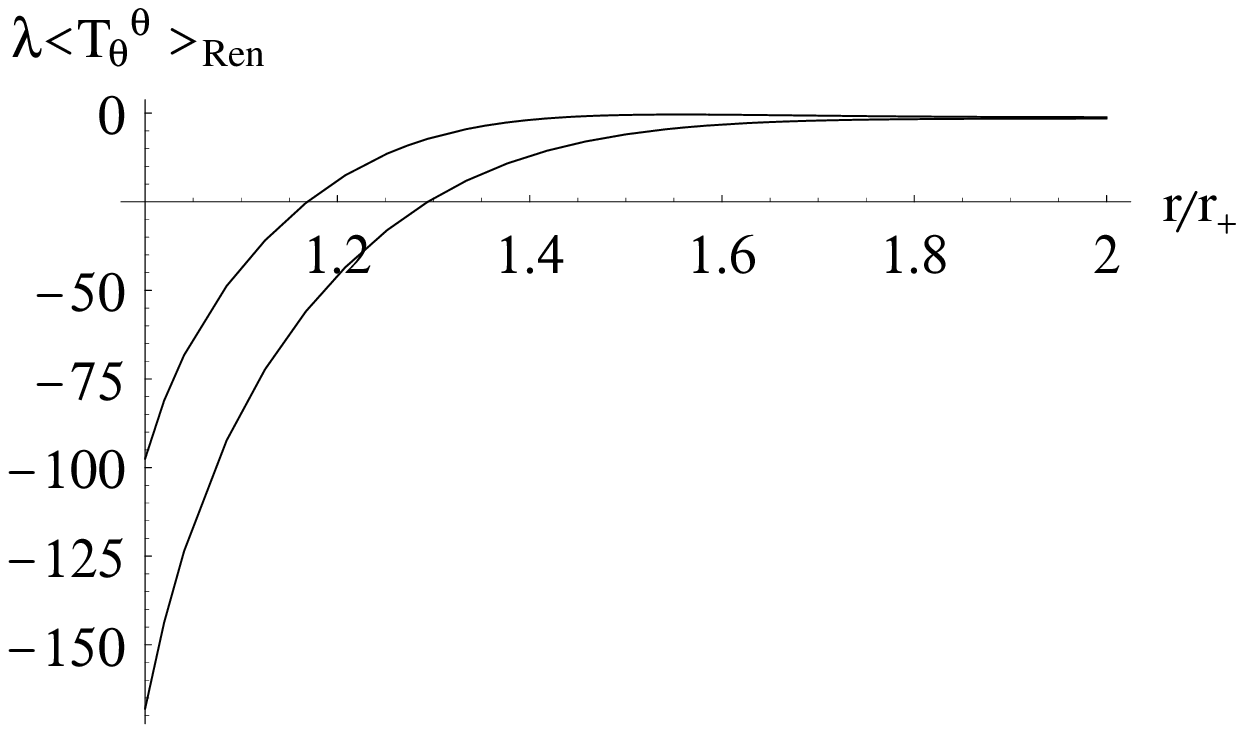}}}
   \vspace{5mm}
   \caption{This graph shows the radial dependence of the rescaled
 component $\langle T_{\theta}^{\theta}\rangle_{ren}$ $[\la = 90 (8 M)^{4}
 m^{2}\pi^{2}]$ of the renormalized stress-energy tensor of the
 massive conformally coupled scalar field. Top to bottom the curves are 
 respectively for the extremal ABG geometry and the 
 extremal RN black hole.} 
   \label{fig10}
   \end{figure}
   %%%%%%%%%%%%%%%%%%%%%%%%%%%%%%%
   %%%%%%%%%%%
   %%%%%%%%%%% TABLES
   %%%%%%%%%%%
 \begin{table}
 \caption{The coefficients $\al_{i}^{(s)}$ for the massive scalar, spinor, and vector
 field}
\begin{tabular}{cccc}
& s = 0 & s = 1/2 & s = 1 \\   \tableline
$\al^{(s)}_{1} $ & $ {1\over2}\xi^{2}\,-\,{1\over 5} \xi $\,+\,${1\over 56}$ & $- {3\over 140}$ &
$- {27\over 280}$\\	  
 $\al^{(s)}_{2}$ & ${1\over 140}$ & ${1\over 14}$ & ${9 \over 28}$\\ 
 $\al^{(s)}_{3}$ &$ \left( {1\over 6} - \xi\right)^{3}$& ${1\over 432}$ &$ - {5\over 72}$\\ 
 $\al^{(s)}_{4}$ & $- {1\over 30}\left( {1\over 6} - \xi\right) $& $- {1\over 90}$ & ${31\over 60}$\\
 
 $\al^{(s)}_{5}$ & ${1\over 30}\left( {1\over 6} - \xi\right)$ &$ - {7\over 720}$ &$ - {1\over 10}$\\
 
 $\al^{(s)}_{6}$ & $ - {8\over 945}  $& $- {25 \over 376}$ & $- {52\over 63}$\\
 
 $\al^{(s)}_{7}$ & ${2 \over 315}$ & $ {47\over 630}$  & $- {19\over 105} $\\	
 $\al^{(s)}_{8}$ & ${1\over 1260}$ & ${19\over 630} $ & ${61\over  140} $\\	  
 $\al^{(s)}_{9}$ & ${17\over 7560}$& ${29\over 3780}$ & $- {67\over 2520}$\\	  
 $\al^{(s)}_{10}$ & $- {1\over 270}$ & $ - {1\over 54} $  & $ {1\over 18}$\\	  
 \end{tabular}
 \label{table1}
 \end{table}
 %%%%%%%%%%
 %%%%%%%%%%
 %%%%%%%%%%
 \begin{table}
\caption{Location of $r_{+}$ and $r_{-}$ of ABG and RN black holes for exemplar
values of $|e|/M.$}
\begin{tabular}{lcccc}
$|e|/M$ & $r_{-}/M$ (ABG)& $r_{+}/M $ (ABG) & $r_{-}/M$ (RN) & $r_{+}/M $  (RN)\\ \tableline 
0.10 & 0.001 &  1.995 & 0.005 & 1.995 \\
0.50 & 0.060 &  1.866 & 0.134 & 1.866 \\
0.95 & 0.422 &  1.356 & 0.688  & 1.312 \\
extremal & 0.871 &  0.871    &  1 & 1 \\
\end{tabular}
\label{table1}
\end{table}


\begin{references}
%
\def\cmp#1#2#3{{ Commun. Math. Phys.} {\bf #1}, #2 (#3)}
\def\lmp#1#2#3{{ Lett. Math. Phys.} {\bf #1}, #2 (#3)}
\def\hpa#1#2#3{{ Hell. Phys. Acta} {\bf #1}, #2 (#3)}
\def\grg#1#2#3{{ Gen. Rel. Grav.} {\bf #1}, #2 (#3)}
\def\pr#1#2#3{{ Phys. Rev.} {\bf #1}, #2 (#3)}
\def\prl#1#2#3{{ Phys. Rev. Lett.} {\bf #1}, #2 (#3)}
\def\prd#1#2#3{{ Phys. Rev. D} {\bf #1}, #2 (#3)}
\def\pl#1#2#3{{ Phys. Lett} {\bf #1}, #2 (#3)}
\def\pla#1#2#3{{ Phys. Lett. A} {\bf #1}, #2 (#3)}
\def\plb#1#2#3{{ Phys. Lett. B} {\bf #1}, #2 (#3)}
\def\prep#1#2#3{{ Phys. Reports} {\bf #1}, #2 (#3)}
\def\phys#1#2#3{{ Physica} {\bf #1}, #2 (#3)}
\def\jcp#1#2#3{{ J. Comput. Phys.} {\bf #1}, #2 (#3)}
\def\jmp#1#2#3{{ J. Math. Phys.} {\bf #1}, #2 (#3)}
\def\jpm#1#2#3{{ J. Phys. A: Math. Gen.} {\bf #1}, #2 (#3)}
\def\cpr#1#2#3{{ Computer Phys. Rept.} {\bf #1}, #2 (#3)}
\def\cqg#1#2#3{{ Class. Quantum Grav.} {\bf #1}, #2 (#3)}
\def\cma#1#2#3{{ Computers Math. Applic.} {\bf #1}, #2 (#3)}
\def\mc#1#2#3{{ Math. Compt.} {\bf #1}, #2 (#3)}
\def\apj#1#2#3{{ Astrophys. J.} {\bf #1}, #2 (#3)}
\def\apjs#1#2#3{{ Astrophys. J. Suppl.} {\bf #1}, #2 (#3)}
\def\acta#1#2#3{{ Acta Astronomica} {\bf #1}, #2 (#3)}
\def\apl#1#2#3{{Ann. Physik. (Leipzig)} {\bf #1}, #2 (#3)}
\def\sa#1#2#3{{ Sov. Astro.} {\bf #1}, #2 (#3)}
\def\sia#1#2#3{{ SIAM J. Sci. Statist. Comput.} {\bf #1}, #2 (#3)}
\def\aa#1#2#3{{ Astron. Astrophys.} {\bf #1}, #2 (#3)}
\def\mnras#1#2#3{{ Mon. Not. R. astr. Soc.} {\bf #1}, #2 (#3)}
\def\npb#1#2#3{{ Nucl. Phys. B} {\bf #1}, #2 (#3)}
\def\prsla#1#2#3{{ Proc. R. Soc. London, Ser. A} {\bf #1}, #2 (#3)}
\def\app#1#2#3{{Acta Phys. Polon. {\rm B}} {\bf #1}, #2 (#3)}
\def\prog#1#2#3{{Prog. Theor. Phys.} {\bf #1}, #2 (#3)}
%
\def\hepth#1#2{{ hep-th }{\bf #1} (#2)}
\def\grqc#1#2{{ gr-qc }{\bf #1} (#2)}
%
%%%%%%%%%%%%%%%%%%%%%%%%%%%%%%%%%%%%%%%%%%%%%%%%%%%%%%%%%%%%%%%%%%%%%%

\bibitem{DeWitt2}B. S. DeWitt, \prep{19 C}{297}{1975}.
\bibitem{Julian}J. S. Schwinger, Phys. Rev. {\bf 82}, 664 (1951).
\bibitem{frolov1}V. P. Frolov and A. I. Zel'nikov, \plb{115}{372}{1982}.
\bibitem{frolov2}V. P. Frolov and A. I. Zel'nikov, \plb{123}{197}{1983}.
\bibitem{frolov3}V. P. Frolov and A. I. Zel'nikov, \prd{29}{1057}{1984}.
\bibitem{frolov4}V. P. Frolov,	Proceedings of the Lebedev Institute
of the Academy of Science of USSR {\bf 169} (1986).
\bibitem{frolov+novikov}V. P. Frolov and I. D. Novikov,
{\it Black Hole Physics} (Kluwer Dordrecht 1998).
\bibitem{AHS2}P. R. Anderson, W. A. Hiscock, and D. A. Samuel,\prd{51}{4337}{1995}.
\bibitem{AH1}W. A. Hiscock, S. L. Larson, and P. R. Anderson, 
\prd{56}{3571}{1997}.
\bibitem{taylor}B. E. Taylor, W. A. Hiscock, and P. R. Anderson,
\prd{55}{6116}{1997}.
\bibitem{sushkov1}S. V. Sushkov, \prd{62}{064007}{2000}.
\bibitem{sushkov2}S. V. Sushkov, gr-qc/0009028.
\bibitem{a-h1}B. E. Taylor, W. A. Hiscock, and P. R. Anderson, \prd{61}{084021}{2000}.
\bibitem{a-h2}P. R. Anderson, W. A.Hiscok, and B. E. Taylor, \prl{85}{2438}{2000}.
\bibitem{tomimatsu}H. Koyama, Y. Nambu, and A. Tomimatsu, Mod. Phys. Lett. 
A {\bf 15}, 815 (2000).
\bibitem{kocio}J. Matyjasek, \prd{61}{124019}{2000}.
\bibitem{Beato}E. Ay\'on-Beato and A. Garc\'{\i}a, \plb{464}{25}{1999}.
\bibitem{Lambert}R. M. Corless, G. H. Gonnet, D.E.G. Hare, D. J. Jeffrey,
and D. E. Knuth, Adv. Comput. Math. {\bf 5}, 329 (1996).
\bibitem{gilkey1}P. B. Gilkey, J. Diff. Geom, {\bf 10}, 601 (1975).
\bibitem{gilkey2}P. B. Gilkey, Trans. Am. Math. Soc. {\bf 225}, 341 (1977).
\bibitem{avra1}I. G. Avramidi, hep-th/9510140.
\bibitem{avra2}I. G. Avramidi, Teor. Mat. Fiz, {\bf 79}, 219 (1989).
\bibitem{avra3}I. G. Avramidi, Nucl. Phys. {\bf B 355}, 712 (1991).
\bibitem{avra4}I. G. Avramidi, \plb{238}{92}{1990}.

\end{references}
\end{document}